\documentclass[table]{article}
\usepackage[utf8]{inputenc}
\usepackage[russian,english]{babel}
\usepackage[T2A]{fontenc}
\usepackage{tabularx}
\usepackage{pgfplots}
\usepackage{pgfplotstable}
\usepackage{longtable}
\usepackage{pgfplots}
\usepackage{amsmath}
\usepackage{amssymb}
\usepackage{amsfonts}
\usepackage{amsthm}
\usepackage{tikz}
\usepackage{caption}

\title{IT companies: the specifics of social networks valuation}
\author{Yupatova K.V.\footnote{Student of Saint-Petersburg State University, Saint-Petersburg, Russia; kvyupatova@gmail.com}, Malafeyev O.A.\footnote{Doctor of Physics and Mathematics Sciences, Professor, Saint-Petersburg State University, Saint-Petersburg, Russia; malafeyevoa@mail.ru}, Lipatnikov V.S.\footnote{PhD in Economics, Associate Professor, National Research University High School of Economics, Saint-Petersburg, Russia; vitalist@mail.ru}, Bezrukikh V.Y. \footnote{General Director, BaltKotloMash LLC, 57 Sedova Street, Saint-Petersburg, Russia; vub@bkm-spb.ru}}
\date{November 2022}

\usepackage{natbib}
\usepackage{graphicx}
\usepackage{url}

\begin{document}

\maketitle

\section* {Abstract}

The study discusses the main features which affect the IT companies valuation on the example of social networks. The relevance of the chosen topic is due to the fact that people live in the information age now, and information technologies surround us everywhere. Because of this, social networks have become very popular. They assist people to communicate with each other despite of the time and distance. Social networks are also companies that operate in order to generate income therefore their owners need to know how promising and profitable their business is.

The social networks differ from traditional companies in this case the purpose of the research is determining the features of social networks that affect the accuracy and adequacy of the results of company valuation.

The paper reviews the definitions of information technology, social networks, history, types of social networks, distinguishing features based on domestic and foreign literature. There are analyzed methods of assessing the value of Internet companies, their characteristics and methods of application.

There is the six social networks evaluation was assessed in the practical part of the study: Facebook, Twitter, Pinterest, Snapchat, Sina Weibo and Vkontakte on the basis of the literature studied and the methods for evaluating the Internet companies which recommended in it, including the method of discounting the cash flow of the company as part of the income approach and the multiplier method as part of a comparative approach. Based on the analysis, the features that affect the social networks valuation are identified.

\textit{Key words:} information technology, IT companies, social networks, income approach, comparative approach, discounting cash flows, multiplier method, real option method, the features of social networks, social networks valuation, Black Sholes formula, Dattar-Matthews real option valuation method,

\newpage
\tableofcontents

\newpage
\section{Introduction}

For the first time, information and communication technologies were actively discussed in the 1960s, when the first information systems appeared, and society embarked on the development of high technologies. Analysts predicted the need and importance of creation and the development of the World Wide Web for society, so huge amounts of money had been invested in the development of the Internet and information technology by the end of the 1990s. This has led to the fact that information technology has become an important and integral part of human life. All of this contributed not only to the development of the Internet, but also to the development of Internet companies, in particular social networks related to them.

Almost every year new similar companies appear on the social media market that offer their functions and attract more and more new users. The invention of social networks has simplified the social life and opened up new opportunities. Also now, in 2020, due to the unfolding coronavirus pandemic, people are forced to stay at home in order to comply with quarantine and self-isolation in order to prevent the spread of infection, therefore, the Internet and social networks are heavily burdened by increased traffic. Now, as never before, social networks have become in demand because they help people keep in touch at a distance, carry out work processes and manage them, learn new information or just spend their free time.

Since social networks are among fast-growing companies, and their number continues to grow rapidly, and as a result of their competition, the issue of correctly valuation of such companies is very relevant.

It is worth noting that there are not enough studies on this topic, all of them are more related to the study of Internet companies, but since social networks are one of the types of such companies, they also touch on the topic of social networking companies, however, it is devoted a little bit to the analysis of features of social networks valuation in these works.

Given the high attractiveness of social networks for investors, it is necessary to estimate the value of the company in terms of the features which appears due to the industry or the direction in which it operates.

The purpose of this study is to analyze the features of social networks that affect the assessment of their value.

The following tasks were set to achieve the goal:

\begin{itemize}
  \item Identify specific characteristics of social networks;
  \item Explore approaches and methods for assessing the value of Internet companies;
  \item Analyze how the use of these methods takes into account the characteristics of not only Internet companies, but specifically social networks;
  \item Research in practice these methods on the example of social networks;
  \item Conclude based on a comparison of the results obtained by these methods.
\end{itemize}

The subject of the study is the analysis of the features of assessing the value of social networks, which affect the value of the company when evaluating it, and the object is the cost of social networks.

This study includes three chapters. The first chapter discusses the theoretical foundations of the information technology industry, the great attention is paid to Internet companies and social networks: definitions, nature, history of social networks, the current situation on the market of social networks and their features affecting the cost estimate. The second chapter describes the traditional approaches and methods for assessing the value of Internet companies, which according to the analyzed literature sources are most suitable for evaluating such companies. In the third chapter, calculations are performed using the described methodology in the second chapter, based on data of six social networks Facebook, Twitter, Pinterest, Snapchat, Sina Weibo and Vkontakte, the results are compared, their interpretation and conclusions are made regarding the accuracy of evaluating social networks using the selected methods.

Due to the insufficient knowledge of the social network market, the theoretical significance of the study is to identify the features of social networks and their analysis that affect the company valuation of this kind by traditional methods. The practical significance is that the technique used in this work can help potential investors and owners to determine the further attractiveness of the project, and to analysts in the study of the social networking market.
 
\section{Theoretical aspects of evaluating the cost of social networks}
\subsection{The essence, history, content and classification of social networks}

The concept IT-technology stands for “information technologies”. Also for a more capacious designation of information technologies in our language the abbreviation "IT" is often used. At the moment, scientists are giving explanations regarding the interpretation of the concept of IT-technologies, mentioning that this term refers to both information technology processes and the whole set of disciplines and areas of business activity of people. The using of information technology is implemented in order to control and manage data and information, the formation, processing, keeping, ordering, as well as their restriction according to the privacy and security policy.

The boundaries of the information technology activities have been clearly defined by UNESCO, thus approving the definition that information technology is a complex of disciplines that includes science, technology and engineering. The IT sector contributes to maximizing labor efficiency by improving work processes in terms of working conditions for employees who process information, and identifying new approaches to improve task performance. 
 Information technology implies the use of the most modern methods of organizing activities and technical capabilities that contribute to ensuring the interaction between employees and equipment, in other words, IT allows you to combine the practical application of technology and human capabilities.

The field of information technology requires quite a lot of costs causing some difficulties, since when it is introduced into the work activity, it is necessary to make monetary investments not only in the purchase of the appropriate equipment, but also in the training of the appropriate specialists.

The broadest interpretation of the concept IT involves the analysis of areas that perform the following actions with data:

\begin{itemize}
  \item Forming;
  \item Transporting;
  \item Saving;
  \item Perceiving.
\end{itemize}

It is worth highlighting the key features of the functionality of information technology. These include:

\begin{itemize}
  \item Data transmission at any distance;
  \item Structured and standardized exchange of information by using of algorithms;
  \item The use of computer capabilities in order to saving information and make it accessible to end users.
\end{itemize}

Based on this, due to the emergence of the Internet, the information technology market offers a huge number of new developments to simplify a person’s life, therefore, his demand among consumers continues to increase. Products and services offered by the IT segment include the development of equipment, software, as well as, for example, the company's internal systems for interactions between employees and customers or users, and much more.

However, in the 21st century, social networks got a tremendous development, without which it is difficult to imagine life of society, since they allow to interact with people almost anywhere in the world.

Under the social Internet network (abbreviation "social network") is understood as an interactive multi-profile online platform, created as a method of supporting interaction between people. A social network is a website with automated functions that allow users to meet new people in order to communicate and create social connections between them. Also, the function of social networks is entertainment, for example, by listening to music, watching movies, etc. The content of such Internet companies is filled in by the users of this network.

The development of social networks origins to 1978, when the first computerized bulletin board (CBBS), developed by Ward Christens, appeared. The idea of creating this virtual board was based on the bulletin board that hung at IBM, on which employees attached notes and ideas for new products. Thus, the developed electronic board, working on the foundation of a special protocol for transferring files, allowed employees to receive new information and share it with others. This was the beginning of a new stage in the development of information technology in the field of communication. Later, in the early 1980s, this idea was modernized by the Microsoft Disk Operating System, the storage of information in electronic boards became more streamlined and systematized, and their speed increased, allowing to attracted more and more new users. 

With the advent and development of commercial Internet service providers (ISP), the use of a telephone line made it possible to access the Internet, which led to the overflow of virtual boards with participants connected to them, so developers began to think about creating special mass platforms that allowed them to be used as public portals or forums.

A little later, in 1995, Todd Crieselman and Stefan Paterno are students of Cornell created the virtual world “theglobe.com” with the aim of register users with a personal account and communicate with other participants, based on their own hobbies and interests. However the idea never got a big scale, ceasing to exist. 

In 1995, a website Classmates.com was launched, developed by Randal Condras. Initially, being a free service, the main idea of the portal was assistance for graduates to constantly keep in touch with friends, to build sympathy and relationships, or to renew relations with people with whom they had contact in the past. Subsequently, Classmates.com became a paid subscription resource, which served as an innovation for such services. This has contributed to the appearance of dozens of new similar sites, becoming an example of a business model.

The turning point for social networks occurred in 2004, when Facebook founded by Mark Zuckerberg entered the social networking market. Facebook had the name “Thefacebook” and was created for students of Harvard University at the very beginning, but a little later, due to its popularity, it expanded the boundaries of its functioning throughout Boston, and then became accessible to students of any educational institutions in the USA. In 2006, Facebook became available to all Internet users who have email.

It is worth noting that social networks include companies that meet certain criteria. By virtue of this, Duzhnikova identifies the following features characteristic of social networks:

\begin{itemize}
  \item Creating a user’s personal page - a profile with main information containing his name, gender, date of birth, interests, etc., and also, if desired, creating a page with complete anonymity. You can also create closed accounts so that only close friends and subscribers are able to see the information.
  \item Interaction between users of a social network by sending messages, music, files, rating by likes, reposts, comments, etc.
  \item Creating a page to achieve a specific goal - searching for familiar persons, blogging, viewing content, talking about business, meeting people, etc.
  \item Satisfaction the needs from creating a page on a social network, for example, meeting the need for communication. 
\end{itemize}

The development of social networks has led to huge investment attractiveness, so the question of correctly assessing the cost of this type of company is becoming increasingly acute.

Besides to the specific features of social networks, scientists also distinguish their types, subdividing into the following categories:

\begin{itemize}
  \item Massive - social networks the access to which is unlimited because of lacking a specific thematic focus in their using. However, with all the freedom of using this kind of social networks, it is worth noting that their activities should be carried out in accordance with the law and not contradict it.
  \item Photo and video hosting - communication in these social networks occurs to a greater extent through the publication of photos and videos. As a rule, such social networks have the demand among people of creative professions and bloggers in order to get more views, comments and ratings of their work.
  \item Thematic - social networks with a clear focus, for example, social networks that unite users by interests, professions or for the purpose of acquaintance, the so-called dating sites.
\end{itemize}

Specialists do not give a general distribution of social networks by their types, therefore, there are many of its classifications. One of these classifications is offered by the French social media expert Frederic Cavazza in the form of a social media map, which he develops for each year. In 2019, the such card was divided into six categories of services, which are presented in table 1:

\begin{table}[h!]
\begin{center}
\begin{tabularx}{1.0\textwidth} { 
  | >{\raggedright\arraybackslash}X 
  | >{\centering\arraybackslash}X 
  | >{\centering\arraybackslash}X | }
 \hline
\textbf{Name of category} & \textbf{Description} & \textbf{Example of resources} \\
 \hline
\textbf{Publishing} & Resources for publishing texts & Wikipedia, Tumblr, Twitter etc. \\
 \hline
\textbf{Sharing} & Resources that allow you to share content and share information & Deezer, Pinterest, SlideShare etc. \\
 \hline
\textbf{Messaging} & Messaging services, otherwise called messengers & Viber, Skype, Telegram, iMessage, Kik etc. \\
 \hline
\textbf{Discussing} & Discussing resources & Ask.fm, Reddit, Disqus etc. \\
 \hline
\textbf{Collaborating} & Collaboration services, for example, corporate networks for organizations & TalkSpirit, Slack, Hangouts Chat etc. \\
 \hline
\textbf{Networking} & Social networks & Facebook, Tinder, Badoo, Nextdoor etc. \\
\hline
\end{tabularx}
\caption*{Table 1. Social Media Map of Frederic Cavazza (2019)}
\end{center}
\end{table} 

The development of information technology has gradually led to a “boom” in social networks. The planet's population is 7,75 billion people, according to the Digital statistics at the end of 2019 - early 2020, Internet users are 59\% of the population, which is 4,54 billion people. 5,19 billion people have mobile phones, which is 67\% of the world's population, of which 92\% have phones with Internet access, and the statistics are growing every year, thereby increasing involvement in the global web. The use of gadgets contributes to the increase in the number of users of social networks. According to the Statista, the number of users of social networks in 2010 was 0,97 billion, and in 2019 reached 2,95 billion, according to forecasts in 2020, population involvement in social networks will reach a value of 3,08 billion, based on this that there is a stable trend of increasing users.

\begin{center}
\begin{figure}[h!]
\begin{tikzpicture}
\begin{axis}
[
ybar,
nodes near coords,bar width=0.3cm,
symbolic x coords={2010, 2011, 2012, 2013, 2014, 2015, 2016, 2017, 2018, 2019},
enlarge x limits=.1,
]
\addplot [ybar,fill=blue] coordinates{ (2010, 0.97) (2011, 1.22) (2012, 1.4) (2013, 1.59) (2014, 1.91) (2015, 2.14) (2016, 2.28) (2017, 2.48) (2018, 2.78) (2019, 2.95)};
\end{axis}
\end{tikzpicture}
\caption*{Chart 1. The number of social networks users  in the world in 2010-2019 (in billion)}
\end{figure}
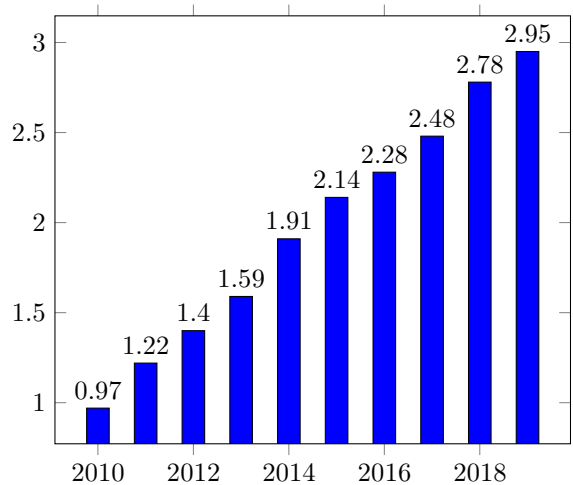
\end{center}

According to the Digital statistics, person spends more than 100 days a year on the Internet an average, which is about 6 hours and 43 minutes a day, for Russians this index is higher and reaches to 7 hours and 17 minutes. The greatest value was recorded among the inhabitants of the Philippines - there people sit on the Internet for 9 hours and 45 minutes.

According to the counts the population uses social networks and messengers almost 50\% of all time spent on the Internet. At the same time, the users of social networks are online in messengers and social networks, not only through mobile applications, but also through a PC, so this indicator for an average user is 2 hours and 45 minutes per day. Similar observations were found among the age group of 16 to 64 years. It is also worth noting that the value varies enough widely across countries, for example, the inhabitants of the Philippines remain leaders among countries with active users of social networks, the spending time is 3 hours 43 minutes in such resources, and the most uninvolved are the Japanese, they spend a day on the activity only 45 minutes. Residents of Russia use social networks including messengers for about 2 hours 26 minutes a day.

\pgfplotsset{width=9cm,height=9cm}
\begin{center}
\begin{figure}[h!]
\begin{tikzpicture}
\begin{axis}[ 
xbar, xmin=0,
xmax = 3000,
xtick=\empty,
xlabel={Number of active users (million)},
symbolic y coords={{Pinterest},{Twitter},{Snapchat},{Kuaishou},{Reddit},{Sina Weibo},{Qzone},{QQ},{Douyin/Tik Tok},{Instagram},{Weixin/WeChat},{Facebook Messenger},{WhatsApp},{YouTube},{Facebook}},
ytick=data,
nodes near coords, nodes near coords align={horizontal},
ytick=data,
]
\addplot [xbar,fill=blue] coordinates {(366,{Pinterest}) (386,{Twitter}) (398,{Snapchat}) (400,{Kuaishou}) (430,{Reddit}) (516,{Sina Weibo}) (517,{Qzone}) (731,{QQ}) (800,{Douyin/Tik Tok}) (1000,{Instagram}) (1265,{Weixin/WeChat}) (1300,{Facebook Messenger}) (2000,{WhatsApp}) (2000,{YouTube}) (2498,{Facebook})};
\end{axis}
\end{tikzpicture} 
\caption*{Chart 2. The most popular social networks by the number of active users in the world as of April 2020 (in million)}
\end{figure}
\end{center}

The graph shows that the Facebook is the most popular social network in the world. In addition, this is the first online platform, which the number of users has exceeded 1 billion and according to statistics at the beginning of 2020 totals 2,498 billion people. Following it are YouTube video hosting and WhatsApp messenger, which have the same number of users in 2 billion. At 7th place is Tik Tok, known in China as Douyin. Since autumn of 2019, the number of its users began to grow in huge sizes reaching the mark of 800 million. This social network has become the main trend of the year with 500 million (62,5\%) of active users of Tik Tok, these are residents of China, which is the birthplace of this online platform. A few other Chinese social networks also have a good reputation in the market and occupy fairly high positions in the ranking, for example, QQ (8th place), Qzone (9th place), Kuaishou (Kwai) (12th place) and Sina Weibo (10th place). It is worth noting that many popular and well-known social networks have analogues in China that are practically not lose to Western social networks, they include WeChat known in China as Weixin, which is an analog of WhatsApp, and now combines the functions of Facebook, Instagram , WhatsApp, Apple/Google Play, and even Uber. And the Chinese social network Sina Weibo (founded in 2009), which serves as an analogue of Twitter (created in 2006), is settled above the western resource at 10 and 14 places respectively. Pinterest was the fastest growing independent social network by the number of unique visitors per month; their number is approximately 10 million. Russian social networks VKontakte and Odnoklassniki remain popular in Russia and in Europe, especially in their region, but their positions have declined in relation to the world, so they occupied 14th and 15th places in the ranking of social networks in 2018, but did not hit in the top 15 most popular social networks in the world to the beginning of 2020.

It is interesting to see the reasons why the social networks are attracted for people, among them are:

\begin{enumerate} 
  \item The use of gadgets by society is growing rapidly.

We can stay in touch anytime and anywhere using mobile devices and the social networks contribute to this through free audio and video calls and text and voice messages once again.

  \item  The need for public acceptance 
  
The social networks have become a platform for self-affirmation and/or for earning for many people. Numerous likes, comments, reposts serve as a tool to increase their own self-esteem and awareness of their own importance for other people, and a large number of followers lead to the emergence of new bloggers who further monetize their audience.

 \item  The need for communication
 
Some people just use social networks to communicate not only with friends, but also to meet new people for relationships, friendships and exchange of information. 

\item  The view the content
 
Most social networks also allow you to view various collections based on the interests of the user, news, listening to music, watching movies, uploading and rating photos, following events in the world and other people's lives.

\item  The convenience of use

On the one hand, the social networks are advanced features that allow you to use them for different purposes; on the other hand, the interface is often limited to a certain functionality that is understandable in use.

\end{enumerate}

Social networks are closely related to business. Entrepreneurs make huge money investments in this industry and they are a key layer of society that actively stimulates its development, thus enabling scientists and researchers to improve ways of working with information, and companies to be most competitive and carry out their activities with the highest possible efficiency.

\subsection{Features of assessing the cost of social networks}

Internet companies, namely social networks have many specific characteristics that distinguish them from traditional companies, for example, the industrial sector. The influence of such characteristics helps to identify features that must be taken into account when evaluating such companies as they can add their own difficulties in the calculations, introducing deviations into them. 

It is worth noting that social networks are fast-growing companies because the number of users is increasing rapidly every year, but despite the fact that the number of involvement is rising, the competition among them is also becoming stronger therefore the need to offer unique products and services for their users is very acute in order to maintain leadership and be in trends.
It is also worth considering the fact that the company is an attractive investment object from which yield will be in the future therefore for investing in the development of a particular type of business it is necessary all risks must be taken into account, and the reliability of the assessment depends on the correctly selected area for evaluating the company valuation.
Social networks differ from traditional companies because they are characterized by certain features that affect the company valuation afterwards therefore the future leaders of such companies need to empirically estimate the value of the social networks taking into account their characteristics. 

\begin{enumerate} 
  \item The volatility of capital structure 

Social networks are companies whose financing is mainly provided by venture investors or their owners. In this case social networks are a fairly risky type of company for investment because, in many cases, the success of social networks depends on the uniqueness of the idea and the monetization of the number of users. Such projects attract venture investors as they are interested in investing in risky companies, because such enterprises have a higher income in the future. In addition, social networks have no costs for the necessary equipment and materials practically, or at least their share is very small.

  \item  Low cost or lack of it
  
Social networks, as a rule, are free resources, but they have advanced functions that require paid subscriptions, but the prices of these services and products are low enough so that anyone who wants it can buy them. In addition, given the fact that the functioning of social networks on gadgets depends on the availability of an Internet connection, and many social networks and messengers offer free calls and messaging functions, their attractiveness is growing increasing the number of users every year.

 \item  The result of the activity is an intangible product or service
 
Social networks refer to companies which the number of intangible assets prevails over tangible assets. The activities of social networks are associated with the provision of telecommunication services therefore, based on this, the main assets of such companies are digital services and products that can be transmitted by telecommunications. Also, a huge part of the assets is occupied by intellectual capital, the brand, the company's business reputation (in other words, goodwill), and the qualifications of specialists. That is, social networks do not need to purchase a large number of equipment, structures, machines and etc. like traditional companies, for example, the industrial sector, because the costs of social networks are very small in terms of tangible assets.

\item  Small or negative earnings in the early years
 
Companies of the information technology sector are characterized by a rather long stage of unprofitability at the first stages, and then a sharp increase in indicators, thus social networks are no exception. Since at the beginning of the launch, social networks need to create free versions or new free test functions in order to promote the resource and be interesting to people and, as a result, attract as many users as possible so that the incomes begin to increase in future. This is due to the fact that IT sector companies have more opportunities for innovation release.

\item  Advertising revenue

The bulk of income generated by social networks gets from advertising. This is due to the fact that now Internet resources are integrating banner and contextual advertising very actively, so traditional advertising methods are losing their significance gradually. However, high traffic to the social network requires for place this kind of advertising.

\item  The presence of a "network effect"

For social networks, revenues depend on the number of users applying it directly, so the customer growth increases the value of the company in a non-linear way. In addition, the cost of advertising and therefore income of social networks depends on the number of users too. The greater the audience’s involvement in a particular social network it means the more expensive the cost of advertising in this social network.

\end{enumerate}

Such features influence the choice of the method for estimating company valuation therefore depending on how these characteristics are taken into account it depends on how much our results deviate from the real valuation and what adjustments and conclusions will be made based on the results.

\section{Methodology for assessing the value of social networks}
The Internet company valuation, as mentioned earlier, is distinguished by its specificity and structure. For this purpose, there will be considered in this chapter the most suitable approaches and methods regarding the assessment of the social networks valuation. Since the social networks market is little studied then it is worthwhile to rely on the work of authors who conducted research in the field of Internet companies for estimating the value of such companies. Based on this, the approaches and methods have been identified that are most suitable for assessing the value of Internet companies. According to scientists, the traditional approaches are suitable for assessing the value of Internet companies within which income, comparative and costly approaches are applied. However, experts are of the opinion that when using traditional approaches it is income and comparative approaches are most suitable for assessing the company valuation, since in order to calculate an estimate of the value of companies by the costly approach it is required the company to have a large number of tangible assets. Insofar as accounting for tangible assets is complicated in assessing the value of social networks, as mentioned earlier, intangible assets prevail in Internet companies, it is impossible to use a costly approach [100].

Some experts also reduce the valuation of Internet companies to certain methods of income and comparative approaches which include the method of discounting cash flow and the method of industry ratios or multipliers respectively. So Koshkina mentions in her work that the real options method is suitable for assessing the value of such companies [72]. 

Thus, special attention was paid to methods for assessing the value of Internet companies during the analyzing articles on the valuation of companies. Based on this, aside from traditional methods for assessing the company valuation, it will emphasis precisely on methods that are suitable for Internet companies, taking into account the characteristics of social networks as a subject of cost estimation, which include discounting cash flow as part of an income approach and the method of industry ratios or multipliers method as part of a comparative approach.

Internet companies are classified as fast-growing. And investors expect a higher rate of return on investments in financial assets. Based on these thoughts, many authors argue in favor of valuing Internet companies with income approach. Asvat Damodoran also claims that this approach provides more accurate estimates [100]. 

The main idea of income approach is that investments should be made in the company development from investors, who acquire not a set of assets in the form of movable and immovable property in this manner, but income flows which further contribute to the return on investment.

One of the approaches to assessing the value of Internet companies is a comparative approach. For applying the approach it is necessary to have information about the market prices of peer companies, based on what value of peer companies operating in the market are estimated it is determined the market value of the selected company.

To apply this approach, it is necessary that the information about competing peer companies be reliable, otherwise it is impossible to avoid errors.

To evaluate companies using industry-specific ratios or multiples, EV/S (enterprise value/sales) and EV/R (enterprise value/revenue) multipliers are most often used.

\subsection{Income approach: discounting cash flows}

Most of all, this method of company valuation is suitable for assessing the value of companies producing a potentially perspective product and for start-up projects that have little or no income, especially in the early stages. According to many authors, this method most accurately estimates the Internet companies valuation, consequently, it is interesting to test it for assessing the value of social networks considering their specifics.

The essence of the assessment is to determine the value of the company by summing the free cash flows of future periods, namely the sum of the current values of the expected income flows taking into account risk. Thus, the forecasted future income is reduced to the current value using the discount rate, the dimension of which is calculated based on the weighted average cost of capital [4]. 

Free cash flows of the company divided into two types: free cash flow to firm and free cash flow to equity.

To get started, it is considered the first view:
The company's free cash flow stands for Free Cash Flow to Firm, reducing the designation to FCFF or FCF. This method is based on forecasting the cash flows of the evaluated company using a discount rate, which in this case is the weighted average cost of capital. The calculation by this method of discounting cash flows takes into account the cost of equity and borrowed capital of the company.

The formation of cash flow by this method occurs at the expense of the company’s assets therefore it allows determining the cash balance after investing in capital assets.

It is worth noting that company valuation by discounting free cash flows to firm has significance to investors and lenders.

The formula for calculating free cash flow by this way looks as follow:

\begin{equation}
FCFF=EBIT\ast (1-Tax)+DA \pm \bigtriangleup NWC \pm CaPex,
\end{equation}

Where:

\begin{itemize}
  \item EBIT - earnings before tax and interest;
  \item Tax - corporate tax rate;
  \item DA (Depreciation and Amortization) - depreciation and/or amortization of tangible and intangible assets;
  \item $\bigtriangleup$ NWC - change in working capital;
  \item CaPex - Capital Expenditure.
\end{itemize}

The discounting free cash flow to firm takes place according to the following formula using the weighted average cost of capital:

\begin{equation}
WACC= R_d*W_d*(1-T)+R_e*W_e,
\end{equation}

Where:

\begin{itemize}
  \item WACC (Weighted Average Cost of Capital)- weighted average cost of capital (discount rate);
  \item $R_{d}$   - debt rate;
  \item $W_{d}$ - share of debt obligations;
  \item $(1-T)$ - corporate tax rate;
  \item $R_{e}$ - cost of equity;
  \item $W_{e}$ - share of equity.
\end{itemize}

In this case, the WACC describes a situation in which the efficiency of investments to a project or enterprise is higher than the efficiency of an existing business.
When calculating this method, the value of the company is determined by using the terminal value of the company:

Due to the confusion that arises when using this formula, it is worth noting that indicators of changes in working capital and capital investments are mainly taken with negative signs, because it is cost.

\begin{equation}\label{eq:fourierrow}
P_{comp} = \sum \limits_{n=1}^{t=n} \frac{FCFF_t}{(1+WACC)^t} + \frac{TV_n}{(1+WACC)^n}, 
\end{equation}

where 
    
\begin{itemize}
  \item $P_{comp}$ - the value of the company;
  \item $t$ - the number of time period on which the assessment occurs;
  \item $n$ - the total number of periods for calculating cash flows. As a rule, 5 years’ time period or more is taken for evaluation.
  \item $TV_{n}$ (terminal value) - value of the company post-forecast period, coerced to the current time taking into account $g_{n}$ (growth rate after the forecast period). It is calculated by the formula:
\end{itemize}
    
\begin{equation}\label{eq:fourierrow}
TV_{n} = \frac{FCFF_{n+1}}{WACC-g_n}.
\end{equation}   
   
Now consider the second form of discounting cash flow. This method consists in discounting Free Cash Flow to Equity. The accepted designation is FCFE. Discounting of this type of flow is usually popular among company shareholders and owners because it is used to calculate the shareholder value of the company.

FCFE is determined by the following formula:

\begin{equation}
FCFE=NI+DA \pm CaPex \pm \bigtriangleup NWC+Net borrowing
\end{equation}

Where:

\begin{itemize}
  \item NI - net income of the enterprise;
  \item Net borrowing - a value equals to the difference between repaid and received loans.
\end{itemize}

Calculation of the company valuation by free cash flow to equity (FCFE) is used when applying the model of financial asset pricing CAPM (Capital Asset Pricing Model) - Sharpe model. When calculating in this way the discount rate $r_{e}$ is the expected rate of return on a long-term asset, which is calculated by the formula:

\begin{equation}
r_e=r_f+ \beta (r_m-r_f)
\end{equation}

Where:

\begin{itemize}
  \item $r_{f}$ - risk-free rate of return;
  \item $r_{m}$ - expected return on the market portfolio;
  \item ($r_{m}$-$r_{f}$ ) - risk premium for investments in shares equaled to the difference in market rates and risk-free rate of returns;
  \item $\beta$ - coefficient for the market, showing riskiness. Calculated by the formula:
\end{itemize}

\begin{equation}
\beta = \frac{cov(r_{e},r_{m})}{\sigma^2(r_{m})}
\end{equation} 

To calculate the enterprise value it is necessary to calculate the amount of equity according to the following formula:

\begin{equation}\label{eq:fourierrow}
Equity = \sum \limits_{t=1}^{t=n} \frac{FCFE_t}{(1+r_e)^t} + \frac{TV_n}{(1+r_e)^n}
\end{equation}

where $TV_{n}$ is the terminal value

The company valuation is calculated as the sum of the market value of the debt and equity.
In a situation where the appraiser predicts firm bankruptcy with the further sale of its property, for this purpose it is necessary to use the following formula:

\begin{equation}
P_{comp}=(1-L_{ur}) \ast (A_{sum}-O)-P_{liq}
\end{equation}

where 

\begin{itemize}
  \item $P_{comp}$ - the value of the company;
  \item $P_{liq}$ - the cost for the liquidation of the company (insurance, appraiser services, taxes, employee benefits and management expenses);
  \item O - the amount of obligations;
  \item $L_{ur}$ - a discount that is provided in connection with the urgency of liquidation of the enterprise;
  \item $A_{sum}$ - the total value of all company assets after their reassessment.
\end{itemize}

Also it should be noted that when using this formula, the assessment results are affected by the location of the company, the quality of its assets and the situation in the market as a whole [4]. 

The discounting cash flow method is considered the most theoretically reasonable method applied to company valuation. The main advantage of this method is that it takes into account the prospect of further development of the market and the company itself.

\subsection{Comparative Approach: Multiplier Method}

Calculation by this method takes into account the recommended ratios between the selling price of the company’s business and its production and financial indicators. Multipliers are a very simple method to assess the company valuation based on the available data of similar companies. The method most widely used in countries with developed market economies.

This method of valuing companies has several advantages:

\begin{itemize}
  \item Provides the most accurate assessment based on the information used about competitors;
  \item Allows you to pre-think the basis for the implementation of the method;
  \item Applying statistics and modeling by using a PC for the purpose of counting.
\end{itemize}

However, even though this method is one of the most suitable for assessing the value of Internet companies, the probability of obtaining false results or errors remains because it depends on the company strategies [4]. 

The multiplier method is still relatively rarely used due to the lack of sufficient statistics in Russian practice.

To assess the value of the company using the multiplier method, taking into account the characteristics of Internet companies, namely social networks, the following set of multipliers are considered most suitable:

EV/EBIT - where EV is the enterprise valuation, and EBIT is the company's profit after taxes and interest. The advantage of this multiplier is that it takes into account depreciation costs.

EV/EBITDA - where is EBITDA (company profit before interest, taxes and depreciation). The advantage of this multiplier is that they can compare companies based in different countries, since the results of its calculations are not affected by the tax and debt burden.

EV/R - The multiplier of the ratio of enterprise value to revenue, it demonstrates the company's ability to generate revenue and income. It is suitable for fast-growing companies operating at break even.

Also, Drogovoz and Kognovitsky in their article considered "sector" multipliers that are used to assess the cost of Internet resources, taking into account the characteristics of such companies. According to their classification, the multiplier of the ratio of company valuation to the number of active users or subscribers will be relevant in this case, noting that advertising is the main source of income for social networks.

EV/N or EV/S is a multiplier of the ratio of the company's value to the number of unique visitors (N) or subscribers (S) of a social network. In essence, these multipliers are identical with the only difference being that N is an indicator of unique visitors, showing statistics on the number of people who viewed a page on a social network. This is due to the fact that social networks are actively integrating banner advertising, respectively, for them the number of people plays a paramount role. At the same time, S is an indicator of the number of subscribers of a social network who, during registration, filled in personal data, hobbies and interests. This allows developers to display ads based on the preferences of subscribers to this social network. This trend is called targeting. 

In this work, we will use the EV/MAU and EV/DAU multipliers, which demonstrate the ratio of the company's value to the number of active users per month and the ratio of cost to the number of active users per day.

Active are users who can be identified by login, e-mail or ID. In most cases, these are users who have created an account on a social network and who need to log in to it for any interaction on the network.

\subsection{Real Options Method}

The valuation of the company by the method of real options began to be used relatively recently, therefore this approach is very new. Unlike traditional approaches, this method allows you to take into account the uncertainty in the market and the rapidly changing environment, thereby reducing the riskiness of investments. Based on this, the fair value of the company is calculated.

Internet companies have a number of features, including the possibility of their future growth with the instability of cash flows, revenues and costs. Based on this, Internet companies can be evaluated in terms of the uncertainty of the environment in which they develop.

As part of the real options method, the following calculation models are used:

\begin{itemize}
 \item[$\star$] \textbf {Black Scholes formula}
\end{itemize} 

This formula is used to determine the price of a European option. The idea is that the price of the underlying asset that is traded on the market is indirectly set by the market itself. Thus, due to fluctuations in stock prices, investors need to constantly adjust their investment in stocks.

The Black Scholes formula is widely used, since it can be used to evaluate options for various kinds of assets, for example, bonds, equity, currency, etc.

The formula is as follows:

\begin{equation}
C=SN(d_1 )-Xe^{-r(T-t)}N(d_2),
\end{equation}

where:

\begin{equation}
d_1= \frac {\ln \frac {S} {X}+(r+\frac {1} {2} \sigma^2)(T-t)} {\sigma \sqrt {(T-t)}},
\end{equation}

\begin{equation}
d_2=d_1- \sigma \sqrt {(T-t)}.
\end{equation}

Where:

\begin{itemize}
  \item C is the current option price;
  \item S – the value of the underlying asset;
  \item X – the exercise price of the option;
  \item r – the risk-free rate of return (taken as a percentage);
  \item (T-t) - the time period until the exercise of the option;
  \item $\sigma$ - the volatility of the underlying asset;
  \item N (d) - cumulative distribution function;
  \item $\exp$ - exhibitor
\end{itemize}

\begin{itemize}
 \item[$\star$] \textbf {Binomial options pricing model}
\end{itemize} 	

The essence of the method is that a tree is constructed with decision branches with the probabilities of their occurrence, starting from the moment in the future and gradually approaching the present, thus, the estimated future cash flows are brought to their value at the current time [100]. 

It should be noted that traditional methods of discounting are not suitable for calculating the present value inside the tree, because the risk of the option and the price of assets based on it constantly change over time, therefore, they do not have a constant discount rate. Based on this, the market value of future cash flows should be determined using the option pricing method.

\begin{itemize}
 \item[$\star$] \textbf {Datar-Matthews real options valuation method}
\end{itemize} 

The algorithm for evaluating real options which is based on simulation. It allows you to evaluate real options using scenario-based forecasting of cash flows from the company's operating activities.

This approach is based on changing key indicators that affect the value of the company, using preset probabilities in order to simulate the future value of this company.

Calculated by the formula:

\begin{equation}
C_0 = E_0 [max (S_T e^{-\mu t} - X_T e^{-rt}, 0)],
\end{equation}
%

where:

\begin{itemize}
  \item $C_{0}$ - the value of a real option;
  \item $S_{T}$ – the operating profit at time T (random variable);
  \item $\mu$ – the rate of return;
  \item $X_{T}$  – the strike price;
  \item (T-t) - the time period until the exercise of the option;
  \item $S_{T}$, $X_{T}$  – the random variables.
\end{itemize}

The simulation takes place according to the Monte Carlo method, and the input parameters in this approach are various scenarios of cash flows.

The results obtained by the Datar-Matthews method using the Monte Carlo method can be reduced to the results calculated by the Black-Scholes formula. 

\begin{itemize}
  \item[$\star$] \textbf {Fuzzy payout method for evaluating real options}
\end{itemize}

This method is similar to the Datar-Mathews method because scenario forecasting is used to distribute the net present value. In this case, to predict the value of the company, instead of the probabilities, the theory of fuzzy numbers is used.

In the process of the fuzzy payment method, possible variants of the future value of the company are modeled, with which the values of the real option are determined, and as a result, they are assigned different weights to further bring the valuation to the present moment.

The formula for calculating this approach is as follows:

\begin{equation}
ROV= \frac { \int_{0}^{\infty} A(x)dx} { \int_{-\infty}^{\infty} A(x)dx} \times E(A_+),
\end{equation}

where: 

\begin{itemize}
  \item A(x) - the value of the company;
  \item $E(A_+)$ - the average expected positive change in the value of the company;
  \item $\int_{-\infty}^{\infty} A(x)dx$ - the area below the entire distribution area of the predicted changes in the cost of digging;
  \item $\int_{0}^{\infty} A(x)dx$ - the area below the positive distribution area of the predicted changes in the cost of company. 
\end{itemize}

The result of the calculation is a triangular fuzzy number used in subsequent calculations to distribute the future value of a real option.

\section{Identification of features of the social networks valuation}

In the practical part of the study, the cost of six companies representing social networks was calculated. The selected companies were based in different countries of the world, such as the USA, China and Russia. Also, the companies rank different places in global ranking of social networks. The sample contains both the most popular global networks and networks that are popular in certain countries, due to which they occupy lower positions in the global ranking.

• Facebook – it founded in 2004 by Mark Zuckerberg in the USA, the office is located in California. According to the rating for April 2020, it ranks first among the most popular social networks in the world, having 2498 million users.

• Twitter – it founded in 2006 by Jack Dorsey. The head office of Twitter are located in California in San Francisco city, additional offices are in San Antonio and Boston. This Internet resource refers to such type of social networks as a microblog, and its essence is that users can share short notes and messages with their subscribers. Back in 2018, Twitter occupied the first lines of ratings, but according to data for April 2020, it fell to 14th place in the world rating, counting 386 active users.

• Pinterest - the social network was launched in 2010 by Ben Silbermann. The server is situated in San Francisco. The idea of the social network is that users can add posts in the form of thematic collections that can be pinned to the “board”. According to the rating for April 2020, Pinterest is on the 15th place in the top social networks of the world, having 366 million users. Also it is worth noting that due to its design idea the social network is the most popular among women.

• Snapchat – it founded in 2011 by Evan Spiegel, Bobby Murphy and Reggie Brown while studying at Stanford University. The main idea of the Internet resource is the ability to send messages with the attachment of photos and videos, which are stored in correspondence only 24 hours and the deleted. It takes 13th place in the ranking with 398 million users.

• Sina Weibo is a social network launched in 2009 in China. It is a microblog, being a hybrid analogue of Twitter and Facebook because users have the ability to exchange instant messages, post photos or videos. It is distributed more locally - in China, therefore it is the most popular among the inhabitants of the Middle Kingdom. But according to the world rating for April 2020, it ranks 10th among the most popular social networks, reaching the mark of 516 million users.

• VKontakte - the social network founded in 2006 by Pavel Durov in Russia. The main office is located in Saint - Petersburg. This social network is the most popular among residents of Russia and the CIS countries. It has 100 million users.

To carry out calculations of the cost of social networks, the following hypotheses were formulated taking into account their features:

\begin{enumerate} 
  \item When assessing the value of social networks by discounting cash flow of a company, it is necessary to take into account the particular structure of expenses and income of social networks.
  \item  The determination of the value of social networks when applying discounting cash flow can take into account such features as the variability of the capital structure and the volume of advertising.
  \item  “Sector” multipliers EV/DAU and EV/MAU and the average values calculated using all multipliers give the most accurate estimate wit the multiplier method.
  \item  The rating of social networks affects the accuracy of the results obtained by the multiplier method.
\end{enumerate}

\subsection{Calculation of the social network valuations by discounting cash flow of the company}

The discounting cash flow of the firm method was used as part of this approach.

The first step in assessing the value of social networks using this method is to analyze the financial indicators of companies, choose a period for making a cash flow forecast, and determine and justify the expected growth rate. Then there is an assessment and forecast of the cash flows, after which the discount rates are calculated at which the predicted cash flows will be adjusted. The cash flow forecast results are presented in Appendix.

Before explaining the application of the method, it is necessary to identify some parameters that were used in the calculations. A time period of 5 years and a post-forecast period are taken for the forecast period, since in the period it is possible to get appropriate results, having financial data for 9-11 years. To calculate the terminal value of the company, that is, the value in the post-forecast period, it is necessary to use growth rates that do not exceed the general growth rate of the nation economy in which the company operates. Accordingly, the average value for 10 years, actual historical data from 2015 to 2019 and forecast values for 2020-2024 were used as growth rates. As a result, the growth rates were: USA - 2.1\%, China - 6.1\% and Russia 1.22\%. These growth rates were used to calculate the terminal value during the estimation, since they match to the ideas about the future growth rate of the companies. The following corporate income tax values were used in the calculations: USA - 21\%, China - 25\%, Russia - 20\%, relevant at the time of the estimation.

The indicators of companies over the past years was analyzed during the practical part, thus, it was determined that these methods will be used to estimate valuations of four companies: Facebook, Twitter, Sina Weibo and Vkontakte. The impracticability of the valuation of Pinterest and Snapchat is due to the fact that Snapchat and Pinterest started publishing reports only in 2016 and 2017, respectively, and historical data for 3-4 years is not enough to make a forecast for five years. Moreover, both companies at the time of 2019 have a negative indicator of net profit, due to the high costs of research and development. In this way, adequate results of estimating the cost of these social networks using the cash flow method are not able to obtain at the moment.

When estimating, the forecast values of the financial indicators of companies, they were calculated by various methods, depending on how fast and structure the data changed over the historical period of time, and also depending on their size. Basically, the forecasts were built using historical or average growth rates. Also the Excel functions “Trend” and “Forecast Sheet” were used during the making the forecasts.

Data from the financial statements of companies are presented in appendices. To calculate the valuation of social networks, the company's revenue and expenses were predicted. It is worth paying attention to the fact that all of the considered social networks account for a large share of revenue from advertising revenue. Moreover, the share of such income increases in total revenue over time. This trend is shown in Chart 3. Thus, this feature was taken into account during the forecasting of the company's revenue. This is due to the fact that over time the number of users of a social network increases, which means that a much larger number of active users view advertisement, therefore the number of offered ads and clicks on it also begin to grow. As a result, revenue from advertising is also increasing. There is an opinion that the more a social network has users or subscribers, the more expensive advertising in the social net because of its attractiveness. In this case, when calculating the valuation of a social network by the method, one more feature is taken into account - the presence of a “network effect”. It is also worth noting that for all companies, the main items of expenditure are research and development, sales and marketing, as well as “cost price” including various office expenses, electricity, depreciation, purchase of equipment and staff costs. Accordingly, the predicted cash flows and their changes in volume and structure are consistent with historical trends.

\begin{center}
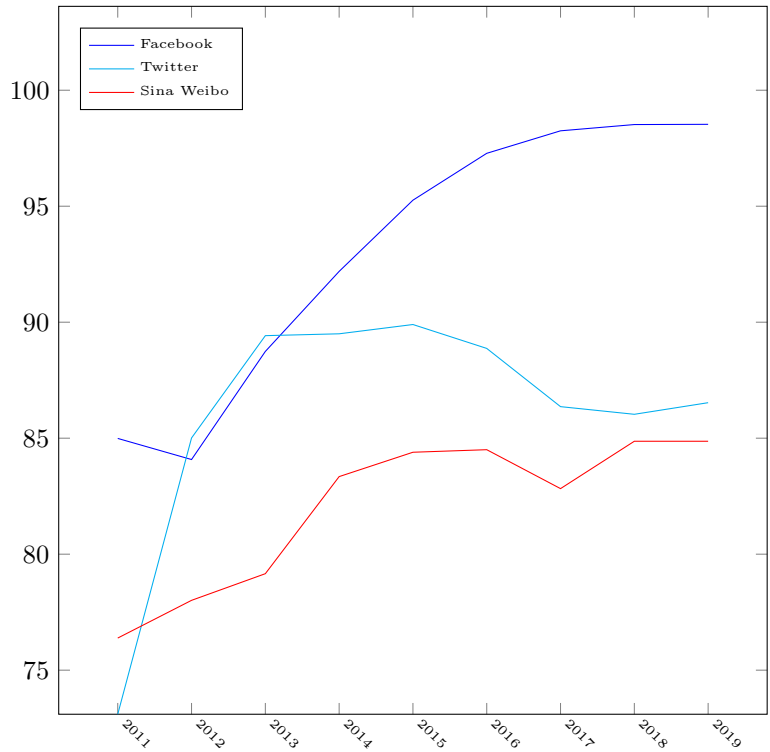
\begin{figure}[h!]
\pgfplotsset{width=11cm,height=11cm}
\pgfplotstableread[col sep=&, header=true]{
description & Facebook & Twitter & Sina Weibo
2011 & 84.99 & 73.10 & 76.384
2012 & 84.08 & 85.01 & 78.010
2013 & 88.74 & 89.42 & 79.159
2014 & 92.19 & 89.50 & 83.344
2015 & 95.26 & 89.90 & 84.394
2016 & 97.28 & 88.87 & 84.504
2017 & 98.25 & 86.36 & 82.826
2018 & 98.52 & 86.03 & 84.868
2019 & 98.53 & 86.53 & 84.868
}\datatableentry
\begin{tikzpicture}
\begin{axis}[
  enlarge y limits ={value=0.2,upper},
  xtick=data,
  xticklabels ={2011,2012,2013,2014,2015,2016,2017,2018,2019},  
  x tick label style={rotate=-45,anchor=west,font=\tiny},
  legend style={font=\tiny,legend pos=north west,legend cell align=left},
]
\addlegendentry{Facebook};
\addplot [color=blue] table [y=Facebook, x expr=\coordindex] {\datatableentry};
\addlegendentry{Twitter};
\addplot [color=cyan] table [y=Twitter, x expr=\coordindex] {\datatableentry};
\addlegendentry{Sina Weibo};
\addplot [color=red] table [y=Sina Weibo, x expr=\coordindex] {\datatableentry};
\end{axis}
\end{tikzpicture}
\caption*{Chart 3. Growth in the share of advertising in total revenue}
\end{figure}
\end{center}

With the aim of company valuations, the cash flows of the social networks were calculated for each forecast period for 5 years, and then discounted using the weighted average cost of capital WACC. Table 2 is showing the data used to calculate the weighted average cost of capital, as well the obtained values for each of the companies.

\begin{table}[h!]
\begin{center}
\begin{tabularx}{1.0\textwidth} { 
  | >{\raggedright\arraybackslash}X 
  | >{\centering\arraybackslash}X 
  | >{\centering\arraybackslash}X
  | >{\centering\arraybackslash}X
  | >{\centering\arraybackslash}X | }
 \hline
\textbf{Company} & \textbf{Facebook} & \textbf{Twitter} & \textbf{Sina Weibo} & \textbf{VKontakte} \\
 \hline
\textbf{Country} & \textbf{USA} & \textbf{USA} & \textbf{China} & \textbf{Russia} \\
 \hline
\textbf{Risk-free rate} & 1,92\% & 1,92\% & 3,18\% & 6,24\% \\
 \hline
\textbf{Risk premium} & 5,20\% & 5,20\% & 5,89\% & 7,37\% \\
 \hline
\textbf{$R_{e}$} & \textbf{7,95\%} & \textbf{5,98\%} & \textbf{9,60\%} & \textbf{17,70\%} \\
 \hline
\textbf{$R_{d}$} & 4,06\% & 4,06\% & 4,35\% & 10,93\% \\
 \hline
\textbf{$\beta$} & 1,16 & 0,78 & 1,09 & 1,56 \\
 \hline
\textbf{E/V} & 84,75\% & 70,76\% & 66,88\% & 71,12\% \\
 \hline
\textbf{E/D} & 15,25\% & 29,24\% & 33,12\% & 28,88\% \\
 \hline
\textbf{Corporate tax} & 21\% & 21\% & 25\% & 20\% \\
 \hline
\textbf{WACC} & \textbf{7,2\%} & \textbf{5,2\%} & \textbf{7,5\%} & \textbf{15,1\%} \\
\hline
\end{tabularx}
\caption*{Table 2. Indicators for calculating WACC}
\end{center}
\end{table} 

Further, the estimation of capital structure ratios were made by based on the available historical data of the companies about the amount of equity and borrowed capital. For this purpose, the average interest rate on loans of each country, in which the selected social networks are based, was taken as the borrowed capital rate ($R_{d}$). When calculating the cost of equity ($R_{e}$), the rate on 10-year government bonds of the USA, China and Russia was used as risk-free rates, as well risk premiums for each of their countries, indicated on the Damodaran website. As the $\beta$  coefficient, it used the values for each of the companies calculated on “Yahoo! Finance” because it is assumed that this coefficient estimate most accurately reflects the sensitivity of the value of the company's assets to market changes. Due to the absence of the $\beta$ coefficient for VKontakte, the average value of $\beta$  coefficients from the Damodaran website was used, which were estimated for Internet companies and advertising campaigns, since social networks receive the main income from advertising. Table 3 presents the values of the forecasted cash flows and the final result of the assessment by the method of discounting cash flow of the company.

\begin{table}[h!]
\begin{center}
\begin{tabularx}{1.0\textwidth} { 
  | >{\raggedright\arraybackslash}X 
  | >{\centering\arraybackslash}X 
  | >{\centering\arraybackslash}X
  | >{\centering\arraybackslash}X
  | >{\centering\arraybackslash}X | }
 \hline
\textbf{Index} & \textbf{Facebook} & \textbf{Twitter} & \textbf{Sina Weibo} & \textbf{VKontakte} \\
 \hline
\textbf{Reduced FCFF-1} & 8 101 381 & 354 912 & (1 581 386) & 126 256 \\
 \hline
\textbf{Reduced FCFF-2} & 14 532 767 & 209 932 & (4 004) & 209 260 \\
 \hline
\textbf{Reduced FCFF-3} & 17 120 689 & 393 403 & 11 732 & 287 881 \\
 \hline
\textbf{Reduced FCFF-4} & 19 225 201 & 590 076 & 26 284 & 382 438 \\
 \hline
\textbf{Reduced FCFF-5} & 20 943 889 & 781 444 & 39 788 & 470 674 \\
 \hline
\textbf{Growth rate in the post-forecast period (economic growth rate)} & 2,1\% & 2,1\% & 6\% & 1,22\% \\
 \hline
\textbf{FCFF for the post-forecast period} & 505 694 337 & 24 962 765 & 4 008 860 & 3 793 508 \\
 \hline
\textbf{Company value, mln.} & \textbf{\$585 618} & \textbf{\$27 293} & \textbf{\$2 501} & \textbf{\$5 270} \\
\hline
\end{tabularx}
\caption*{Table 3. Value of companies, amounts in thousands of US dollars, unless otherwise indicated}
\end{center}
\end{table} 

It is worth noting that the method of discounting cash flow to equity has not been applied for the following reasons.

First of all, there is calculated the total value of the company in this study, and not only just the value of the company for shareholders. The next reason is that to calculate the value by using the discounting cash flow method, it is necessary to make assumptions and forecasts for the debt and the company's future credit policy. When analyzing the financial indicators of social networks, it was detected that they have an unstable financing structure; therefore, it is difficult to make assumptions about the company's future policy regarding debt obligations analytically. Thus, the value of social networks was determined using the cash flow received by both shareholders and lenders.

To verify the effectiveness of the method, a comparison of the results of calculations obtained using the cash flow discounting method with the actual value of the company, estimated by experts as of April 2020, was made.

\begin{table}[h!]
\begin{center}
\begin{tabularx}{1.0\textwidth} { 
  | >{\raggedright\arraybackslash}X 
  | >{\centering\arraybackslash}X 
  | >{\centering\arraybackslash}X
  | >{\centering\arraybackslash}X
  | >{\centering\arraybackslash}X | }
 \hline
\textbf{Company} & \textbf{Estimated value} & \textbf{Actual value} & \textbf{Deviation} \\
 \hline
\textbf{Facebook} & \$585 618 & \$584 350 & 0,22\% \\
 \hline
\textbf{Twitter} & \$27 293 & \$22 072 & 23,65\% \\
 \hline
\textbf{Sina Weibo} & \$2 501 & \$2 171 & 15,21\% \\
 \hline
\textbf{VKontakte} & \$5 270 & \$6 852 & -23,09\% \\
\hline
\end{tabularx}
\caption*{Table 4. Comparison of the results of social network valuations with actual values, amount in millions of US dollars}
\end{center}
\end{table} 

According to the results of the analysis, it can be concluded that the discounting cash flows method of a company is able to estimate the value of social networks taking into account their characteristics, such as the cost of advertising, the variability of the capital structure and the network effect. Deviations for all companies do not exceed 25\%, which can be considered acceptable because the values of companies change regularly depending on the market situation.

\subsection{Calculation of the value of social networks by the multiplier method}

To evaluate social networks by a comparative approach, five different multipliers were used, the choice of which for the companies of this type was justified in the second chapter. Since the multiplier method involves assessing the value of companies based on financial indicators of similar companies, and data about social networks are often not available publicly, thereby the value of six social networks (Facebook, Twitter, Pinterest, Snapchat, Sina Weibo and VKontakte) was estimated by financial results of five other companies from the selected sample. Thus, to estimate each of the companies, the appropriate multipliers for five companies were calculated, the average value for each of the multipliers was taken, and then the result was multiplied by the financial indicator of the analyzed company. It is also worth noting that to calculate the values of companies using the multiplier method, it was used the financial results of 2019, because these data are the most relevant at the time of the study. A summary table of the company results for 2019 used in the calculations is presented in Appendix. Table 5 contains the calculation results of the EV/EBIT, EV/EBIDTA, EV/R, EV/DAU, EV/MAU multipliers for each company.

\begin{table}[h!]
\small
\begin{center}
\begin{tabular}{ | m{6.25em} | m{0.9cm} | m{1.2cm} | m{1.5cm}| m{1.5cm} | m{1.02cm} | m{0.9cm} | } 
  \hline
 \textbf{Multiplier} & \textbf{FB} & \textbf{Twitter} & \textbf{Pinterest} & \textbf{Snapchat} & \textbf{Sina Weibo} & \textbf{VK} \\
 \hline
\textbf{EV/EBIT} & 23,55 & 56,57 & (7,16) & (23,46) & 8,51 & 52,06 \\
 \hline
\textbf{EV/EBIDTA} & 19,13 & 25,79 & (7,31) & (25,62) & 6,88 & 52,06 \\
 \hline
\textbf{EV/R} & 8,27 & 6,38 & 8,52 & 14,13 & 1,00 & 22,93 \\
 \hline
\textbf{EV/DAU} & 352,66 & 145,21 & 38,96 & 111,19 & 9,78 & 297,91 \\
 \hline
\textbf{EV/MAU} & 233,93 & 66,88 & 29,08 & 82,73 & 4,21 & 95,70 \\
  \hline
\end{tabular}
\caption*{Table 5. Multiplier Values}
\end{center}
\end{table}

\normalsize

Further, the value of each of the six companies was determined using the average values of each multiplier for five companies. The results of the calculations can be seen in table 6. If there are negative indications of the multipliers, their module values were applied. Also, the average values of the obtained values were calculated for each social network.

\begin{table}[h!]
\small
\begin{center}
\begin{tabular}{ | m{6.25em} | m{1.2cm} | m{1.2cm} | m{1.5cm}| m{1.5cm} | m{1.02cm} | m{0.9cm} | } 
  \hline
 \textbf{Index} & \textbf{FB} & \textbf{Twitter} & \textbf{Pinterest} & \textbf{Snapchat} & \textbf{Sina Weibo} & \textbf{VK} \\
 \hline
\textbf{EV/EBIT} & 429 372 & 4 175 & 31 907 & 27 596 & 5 184 & 1 527 \\
 \hline
\textbf{EV/EBIDTA} & 316 530 & 7 724 & 20 858 & 18 268 & 4 045 & 497 \\
 \hline
\textbf{EV/R} & 748 940 & 37 951 & 12 047 & 16 162 & 26 055 & 2 289 \\
 \hline
\textbf{EV/DAU} & 199 854 & 24 639 & 45 838 & 36 821 & 42 000 & 3 026 \\
 \hline
\textbf{EV/MAU} & 139 188 & 29 412 & 32 391 & 25 186 & 52 458 & 5 969 \\
\hline
\textbf{Average value} & 366 777 & 20 780 & 28 608 & 24 806 & 25 948 & 2 662 \\
  \hline
\end{tabular}
\caption*{Table 6. Calculation of company values using average multiples}
\end{center}
\end{table}

\normalsize

At the end of the analysis, the values obtained by the multiplier method were compared with the actual cost of each company, which amounted to: Facebook - \$584,350 million, Twitter - \$22,072 million, Pinterest - \$9,741 million, Snapchat - \$24,240 million, Sina Weibo - \$2,171 million, VKontakte - \$6,852 million. Based on this, Table 7 is compiled, which shows the deviations from the values obtained by calculating by each multiplier and the average value, from the actual value of social networks.

\begin{table}[h!]
\small
\begin{center}
\begin{tabular}{ | m{6.25em} | m{1.2cm} | m{1.2cm} | m{1.5cm}| m{1.5cm} | m{1.02cm} | m{0.9cm} | } 
  \hline
 \textbf{Index} & \textbf{FB} & \textbf{Twitter} & \textbf{Pinterest} & \textbf{Snapchat} & \textbf{Sina Weibo} & \textbf{VK} \\
 \hline
\textbf{Actual cost} & 584 350 & 22 072 & 9 741 & 24 240 & 2 171 & 6 852 \\
 \hline
\textbf{EV/EBIT} & \cellcolor{cyan}-27\% & -81\% & 228\% & \cellcolor{green}14\% & 139\% & -78\% \\
 \hline
\textbf{EV/EBIDTA} & \cellcolor{yellow}-46\% & -65\% & 114\% & \cellcolor{cyan}-25\% & 86\% & -93\% \\
 \hline
\textbf{EV/R} & \cellcolor{cyan}28\% & 72\% & \cellcolor{cyan}24\% & \cellcolor{cyan}-33\% & 1100\% & -67\% \\
 \hline
\textbf{EV/DAU} & -66\% & \cellcolor{green}12\% & 371\% & 52\% & 1835\% & -56\% \\
 \hline
\textbf{EV/MAU} & -76\% & \cellcolor{cyan}33\% & 233\% & \cellcolor{green}4\% & 2316\% & \cellcolor{green}-13\% \\
\hline
\textbf{Average value} & \cellcolor{yellow}-37\% & \cellcolor{green}-6\% & 194\% & \cellcolor{green}2\% & 1095\% & -61\% \\
  \hline
\end{tabular}
\caption*{Table 7. Deviations of the company values from actual costs using the multiplier method}
\end{center}
\end{table}

\normalsize

Due to the lack of data on a large number of companies, it was not possible to create a basis for analyzing the value of each company based on a group of social networks that have absolutely similar products registered in one country and also being equally popularity, that is why to determine the best multipliers, it will be expedient to apply results deviating by less than 50\% from the actual value at this stage of the analysis. Based on this premise, the results were divided into three groups:

\begin{enumerate} 
  \item Deviation less than 20\% - the most accurate results highlighted in green color.
  \item Deviation from 21\% to 35\% - medium accuracy results highlighted in blue color.
  \item  Deviation from 36\% to 50\% - the least accurate results, they are highlighted in yellow color. However, it is assumed that these multipliers are able to give a more accurate estimate if the sample for calculating the average value is expanded.
\end{enumerate}

Also, for a detailed analysis, table 8 below shows the positions, which these social networks occupy in the general world ranking of social networks by popularity, and the number of monthly and daily active users.

\begin{table}[h!]
\small
\begin{center}
\begin{tabular}{ | m{4.0em} | m{1.0cm} | m{1.2cm} | m{1.5cm}| m{1.5cm} | m{1.02cm} | m{0.9cm} | } 
  \hline
 \textbf{Index} & \textbf{FB} & \textbf{Twitter} & \textbf{Pinterest} & \textbf{Snapchat} & \textbf{Sina Weibo} & \textbf{VK} \\
 \hline
\textbf{Ranking place} & 1 & 14 & 15 & 13 & 10 & - \\
 \hline
\textbf{DAU} & 1 657 & 152 & 250 & 218 & 222 & 23 \\
 \hline
\textbf{MAU} & 2 498 & 330 & 335 & 293 & 516 & 72 \\
  \hline
\end{tabular}
\caption*{Table 8. The number of active users and position in the global ranking of social networks (million people)}
\end{center}
\end{table}

\normalsize

Based on the results of the analysis, it can be concluded that the multiplier that can most accurately assess the value of social networks is the EV/MAU multiplier, which represents the ratio of the company's value to the number of active users per month. This multiplier is able to take into account the features of social networks, since it estimates the monthly number of active users, and it is used for evaluating such companies specifically. The next good indicator for calculating the cost of social networks is the use of the average value of the company calculated by all multipliers. The use of this method can show the most comprehensive assessment. The third indicator that showed more or less accurate values is the EV/R multiplier, which demonstrates the company's ability to generate revenue. It can be connected with the fact that all social networks have a significant part of the revenue from advertising. Accordingly, exactly these services that create the main value of such companies.

In addition to assessing the quality of the multipliers used, it was also noted that the method showed the least accurate results for Sina Weibo and VKontakte operating in China and Russia. The other four companies are registered in the USA. It is also worth noting that Sina Weibo is more popular in the world than some American social networks and has a larger number of active users. The described factors indicate in aggregate that, in assessing social networks, the country-specific affiliation of social networks plays an important role. As known, there are strict rules regarding the publication of advertisements, as well as drastic restrictive measures imposed on the use of social networks in China. As a result of these measures, Sina Weibo, which operates in China and is more popular than U.S. Twitter, Facebook and Snapchat, has a lower total value. Also, it is suggested that the method could not assess the value of the VKontakte qualitatively due to the fact that the company is much less popular than other analyzed companies. It means that when using the multiplier method to estimate social network valuation, it is worthwhile to select companies that occupy similar positions in the ranking by popularity.

\subsection{Research Results}

In this paper, the methods for assessing the social network valuations were investigated and compared based on all the calculations made.

According to the estimation result, it can be drawn the following conclusions about the features that affect the assessment of the social network valuations:

The discounting cash flow method of the company. Based on the obtained estimates, this method is suitable for assessing the value of social networks, since it estimates their value accurately. Also it is necessary to note that when calculating the cost using this method, it is necessary to predict cash flows, taking into account the fact that it is inherent for social networks to increase the share of advertising revenue. In addition to taking into account the growth in the share of revenue from advertise, this method takes into account such a feature of social networks as the variable structure of capital. In addition, it was found that the main expenses of social networks are research and development, sales and marketing, as well as “cost price”. It is worth considering this fact and maintaining the structure when making expenses forecasts.

The obtained results using the multiplier method show that this method is also appropriate for assessing the social network valuation, but with some clauses. The relatively adequate results of assessing the value of social networks from the calculated multipliers are given by the EV/MAU, EV/R multipliers and the average value for all multipliers. However, among the backgrounds for a future study, it is worth noting that when evaluating social networks using the multiplier method, the selection of companies on the basis of which the analyzed social network will be evaluated should include companies operating in the same country and also enjoying the same popularity.

Thus, based on the above, also it is necessary making conclusions about the hypotheses in this paper.
The first hypothesis regarding the need to take into account the specific structure of income and expenses of social networks when estimating the value by the discounting cash flow method was confirmed. The second hypothesis that the discount method takes into account the characteristics of social networks including the amount of advertising and the variable structure of capital is also confirmed. Besides to the features proposed in the hypotheses, it should be noted that also the method of discounting cash flow to firm (FCFF) showed that the network effect also affects the results of the valuation. The third hypothesis about the accuracy of evaluating the calculation by the multiplier method is refuted partially because though the EV/DAU and EV/MAU multipliers are specific for evaluating social networks, the EV/DAU multiplier cannot give an adequate estimation. However, the EV/MAU, EV/R multiples and the application of the average valuations calculated using all the multipliers will be able to do this, since the latter method covers various indicators of the company. Based on this, the third hypothesis cannot be completely refuted; therefore, it is confirmed partially. The last hypothesis about the impact of the rating on the assessment of the social network valuations using the multiplier method also found its confirmation because in order to use this method it is necessary to make a comparison among close companies in ranking.

However, it is necessary to make a small remark that during the calculations in the framework of this study, it was found that the rating not always can give an accurate description of the market situation, since a company located higher in the global rating is not always more expensive, as calculations showed in practical parts. Sina Weibo ranks 10th in the rating in terms of the number of active users, whose value is 516 million, unlike Snapchat (398 million), Twitter (386 million) and Pinterest (366 million), which occupy 13, 14 and 15 rating accordingly. However, calculations of the company valuations showed that Sina Weibo, in many ways with more active users, is valued much less than all of the other analyzed social networks, including its cost less than those companies that are lower in the rating. Based on this, we can draw another conclusion about the features that affect the value assessment that was identified during the study. The value of a company can be affected by its geographical location, namely, the country which it belongs. In China, the population is much higher than the number of other countries, however, they have a closed Internet and severe restrictions of the activities of social networks, which can cause problems for monetizing subscribers and users, which means generating the income.

\newpage
\section{Conclusion}

In this study, traditional methods for assessing the social network valuations are considered, taking into account the literature studied on the valuation of Internet companies, their features are identified and recommendations for their application are given. In the framework of this paper, calculations were performed according to the data of six social networks Facebook, Twitter, Snapchat, Pinterest, Sina Weibo and VKontakte, and a comparison was made of the results in terms of practice regarding the approaches studied.

In this work, the backgrounds were identified that contribute to the growth of social networking popularity and demand. These include: the use of gadgets by society is growing rapidly, the need for public acceptance, the need for communication, viewing content, ease of use.

Also, in the research with considering the calculations, traditional approaches to assessing the value of social networks are analyzed and it is demonstrated that when choosing a method, it is necessary to take into account factors such as the stability of the financing structure and the market scale. Based on this, the method of discounting cash flow to equity (FCFE), as one of the methods of the income approach, is not appropriate for companies with an unstable financial structure, since it is impossible to make forecasts about the company's future credit policy. The method of discounting cash flow to firm (FCFF) estimates the social network valuations the most accurately, but for this it is necessary to take into account the specific structure of the company's income and expenses. The multiplier method in the framework of the comparative approach should take into account the social network market scale for more accurate assessment because not all multipliers can give adequate assessment results with a small sample.

Further, the features that have the greatest impact on assessing the social network valuations include the variability of the capital structure, the network effect, the position that the social network occupies in the global rating, and the main feature is that such companies receive the main income from advertising. Another feature found in the research is the country where the social network was founded and the location where social networks operate.

The obtained results in the course of this work can be useful for further research of methods for assessing the social network valuations because they are important for practice, since they take into account the identified features of social networks that impact on their value estimation. Due to the limited historical data on companies, the models have small calculation errors (deviations), however the studied methods have shown their advantages and disadvantages and what nuances are worth paying to attention. Therefore, in the future it may become a background for further research on this topic and the generation of new models based, possibly, even on combining the using methods in a hybrid approach.


\newpage

\section{References}

1.	Irina Zaitseva, Oleg Malafeyev, Yuliya Marenchuk, Dmitry Kolesov and Svetlana Bogdanova, Competitive Mechanism For The Distribution Of Labor Resources In The Transport Objective, Journal of Physics: Conference Series, Volume 1172, International Conference on Applied Physics, Power and Material Science 5–6 December 2018, Secunderabad, Telangana, India, \newline
https://doi.org/10.1088/1742-6596/1172/1/012089 

2.	Neverova E.G., Malafeyef O.A., A model of interaction between anticorruption authority and corruption groups, In proc.: AIP Conference Proceedings. 2015. С. 450012.  

3.	Malafeyev O., Awasthi A., Zaitseva I., Rezenkov D., Bogdanova S., A dynamic model of functioning of a bank, In proc.: AIP Conference Proceedings. International Conference on Electrical, Electronics, Materials and Applied Science. 2018. С. 020042.

4.	Kolokoltsov V.N., Malafeyev O.A., Understanding game theory: introduction to the analysis of many agent systems with competition and cooperation, В книге: Understanding Game Theory: Introduction to the Analysis of Many Agent Systems with Competition and Cooperation. 2010. С. 1-286. 
	
5.	Zaitseva I., Malafeev O., Strekopytov S., Bondarenko G., Lovyannikov D., Mathematical model of regional economy development by the final result of labour resources, In proc.: AIP Conference Proceedings. International Conference on Electrical, Electronics, Materials and Applied Science. 2018. P. 020011. 

6.	Malafeyev O.A., Nemnyugin S.A., Ivaniukovich G.A., Stohastic models of social-economic dynamics, In proc.: 2015 International Conference "Stability and Control Processes" in Memory of V.I. Zubov (SCP). 2015. P. 483-485.

7.	Malafeev O.A., Redinskih N.D., Gerchiu A.L. Optimizacionnaya model' razmeshcheniya korrupcionerov v seti. V knige: Stroitel'stvo i ekspluataciya energoeffektivnyh zdanij (teoriya i praktika s uchetom korrupcionnogo faktora) (Passivehouse).  Kolchedancev L.M., Legalov I.N., Bad'in G.M., Malafeev O.A., Aleksandrov E.E., Gerchiu A.L., Vasil'ev YU.G. Kollektivnaya monografiya. Borovichi, 2015. S. 128-140.

8.	Malafeev O.A., Koroleva O.A., Vasil'ev YU.G. Kompromissnoe reshenie v aukcione pervoj ceny s korrumpirovannym aukcionistom, V knige: Stroitel'stvo i ekspluataciya energoeffektivnyh zdanij (teoriya i praktika s uchetom korrupcionnogo faktora) (Passivehouse).  Kolchedancev L.M., Legalov I.N., Bad'in G.M., Malafeev O.A., Aleksandrov E.E., Gerchiu A.L., Vasil'ev YU.G. Kollektivnaya monografiya. Borovichi, 2015. S. 119-127.

9.	Malafeev O.A., Koroleva O.A., Neverova E.G., Model' aukciona pervoj ceny s vozmozhnoj korrupciej, V knige: Vvedenie v modelirovanie korrupcionnyh sistem i processov.  Malafeev O.A. i dr. Kollektivnaya monografiya. Pod obshchej redakciej d.f.-m.n., professora O. A. Malafeeva. Stavropol', 2016. S. 96-102. 
	
10.	Drozdov G.D., Malafeev O.A., Modelirovanie mnogoagentnogo vzaimodejstviya processov strahovaniya, monografiya / G. D. Drozdov, O. A. Malafeev ; M-vo obrazovaniya i nauki Rossijskoj Federacii, Sankt-Peterburgskij gos. un-t servisa i ekonomiki. Sankt-Peterburg, 2010.Malafeyev O., Saifullina D., Ivaniukovich G., Marakhov V., Zaytseva I., The model of multi-agent interaction in a transportation problem with a corruption component, In proc.: AIP Conference Proceedings. 2017. С. 170015. 

11.	Malafeyev O., Farvazov K., Zenovich O., Zaitseva I., Kostyukov K., Svechinskaya T., Geopolitical model of investment power station construction project implementation, V sbornike: AIP Conference Proceedings. International Conference on Electrical, Electronics, Materials and Applied Science. 2018. S. 020066. 	

12.	Malafeev O.A., Nemnyugin S.A., Stohasticheskaya model' social'no-ekonomicheskoj dinamiki, V sbornike: Ustojchivost' i processy upravleniya. Materialy III mezhdunarodnoj konferencii. 2015. S. 433-434.

13.	Zaitseva I., Poddubnaya N., Malafeyev O., Vanina A., Novikova E., Solving a dynamic assignment problem in the socio-economic system, In proc.: Journal of Physics: Conference Series. 2019. С. 012092. 

14.	Malafeyev O., Zaitseva I., Pavlov I., Shulga A., Sychev S., Badin G. company life cycle model: the influence of interior and exterior factors. In proc.: AIP Conference Proceedings. Сер. "International Conference on Numerical Analysis and Applied Mathematics, ICNAAM 2019" 2020. С. 420027.

15.	Malafeyev O.A., Rylow D., Pichugin Y.A., Zaitseva I., A statistical method for corrupt agents detection, In proc.: AIP Conference Proceedings. Сер. "International Conference of Numerical Analysis and Applied Mathematics, ICNAAM 2017" 2018. С. 100014.

16.	Malafeyev O.A., Rylow D., Zaitseva I., Ermakova A., Shlaev D., Multistage voting model with alternative elimination, In proc.: AIP Conference Proceedings. Сер. "International Conference of Numerical Analysis and Applied Mathematics, ICNAAM 2017" 2018. С. 100012.

17.	Malafeyev O.A., Redinskikh N.D., Nemnyugin S.A., Kolesin I.D., Zaitseva I.V., The optimization problem of preventive equipment repair planning, In proc.: AIP Conference Proceedings. Сер. "International Conference of Numerical Analysis and Applied Mathematics, ICNAAM 2017" 2018. С. 100013.

18.	Kolesin I., Malafeyev O., Andreeva M., Ivanukovich G., Corruption: taking into account the psychological mimicry of officials, In proc.: AIP Conference Proceedings. 2017. С. 170014.

19.	Malafeyev O., Rylow D., Novozhilova L., Zaitseva I., Popova M., Zelenkovskii P., Game-theoretic model of dispersed material drying process, In proc.: AIP Conference Proceedings. Сер. "International Conference on Functional Materials, Characterization, Solid State Physics, Power, Thermal and Combustion Energy, FCSPTC 2017" 2017. С. 020063.  

20.	Malafeyev O., Lakhina J., Redinskikh N., Smirnova T., Smirnov N., Zaitseva I., A mathematical model of production facilities location, In proc.: Journal of Physics: Conference Series. 2019. С. 012090. 
	 
21.	Zaitseva I., Ermakova A., Shlaev D., Malafeyev O., Strekopytov S., Game-theoretical model of labour force training, Journal of Theoretical and Applied Information Technology. 2018. Т. 96. № 4. С. 978-983. 

22.	Malafeev O.A., Ahmadyshina A.R., Demidova D.A., Model' tendera na rynke rielterskih uslug s uchetom korrupcii, V knige: Stroitel'stvo i ekspluataciya energoeffektivnyh zdanij (teoriya i praktika s uchetom korrupcionnogo faktora) (Passivehouse).  Kolchedancev L.M., Legalov I.N., Bad'in G.M., Malafeev O.A., Aleksandrov E.E., Gerchiu A.L., Vasil'ev YU.G. Kollektivnaya monografiya. Borovichi, 2015. S. 161-168.

23.	Malafeev O.A., Redinskih N.D., Stohasticheskoe ocenivanie i prognoz effektivnosti razvitiya firmy v usloviyah korrupcionnogo vozdejstviya, V sbornike: Ustojchivost' i processy upravleniya. Materialy III mezhdunarodnoj konferencii. 2015. S. 437-438. 

24.	Titarenko M.L., Ivashov L.G., Kefeli I.F., Malafeev O.A., i dr. Evrazijskaya duga nestabil'nosti i problemy regional'noj bezopasnosti ot Vostochnoj Azii do Severnoj Afriki, Kollektivnaya monografiya / Sankt-Peterburg, Studiya NP-Print, 2013, 576 s.

25.	Malafeev O.A., The existence of situations of $\epsilon$-equilibrium in dynamic games with dependent movements, USSR Computational Mathematics and Mathematical Physics. 1974. T. 14. № 1. S. 88-99. 

26.	Malafeev O.A., Kolesin I.D., Kolokol'cov V.N.. i dr. Vvedenie v modelirovanie korrupcionnyh sistem i processov, kollektivnaya monografiya / pod obshchej redakciej d.f. - m.n. , professora O. A. Malafeeva. Stavropol', Izdatel'skij dom "Tesera",  2016. Tom 1, 224 s.	

27.	Malafeyev O., Kupinskaya A., Awasthi A., Kambekar K.S., Random walks and market efficiency in chinese and indian markets, Statistics, Optimization and Information Computing. 2019. T. 7. № 1. S. 1-25. 

28.	Zaitseva I., Dolgopolova A., Zhukova V., Malafeyev O., Vorokhobina Y., numerical methof for computing equilibria in economic system models with labor force, V sbornike: AIP Conference Proceedings. 2019. S. 450060.

29.	Malafeev O.A. On the existence of Nash equilibria in a noncooperative n-person game with measures as coefficients, Communications in Applied Mathematics and Computational Science. 1995. T. 5. № 4. S. 689-701. 

30.	Malafeev O.A., Redinskih N.D., Smirnova T.E., Setevaya model' investirovaniya proektov s korrupciej, Processy upravleniya i ustojchivost'. 2015. T. 2. № 1. S. 659-664. 

31.	Kirjanen A.I., Malafeyev O.A., Redinskikh N.D., Developing industries in cooperative interaction: equilibrium and stability in process with lag. Statistics, Optimization and Information Computing. 2017. T. 5. № 4. S. 341-347.

32.	Troeva M.S., Malafeev O.A., Ravnovesie v diffuzionnoj konfliktnoj modeli ekonomiki so mnogimi uchastnikami, V sbornike: Dinamika, optimizaciya, upravlenie. Ser. "Voprosy mekhaniki i processov upravleniya" Sankt-Peterburg, 2004. S. 146-153.

33.	Pichugin YU.A., Malafeev O.A., Ob ocenke riska bankrotstva firmy, V knige: Dinamicheskie sistemy: ustojchivost', upravlenie, optimizaciya. Tezisy dokladov. 2013. S. 204-206.

34.	Malafeev O.A., Redinskih N.D., Alferov G.V., Smirnova T.E., Korrupciya v modelyah aukciona pervoj ceny, V sbornike: Upravlenie v morskih i aerokosmicheskih sistemah (UMAS-2014). 7-ya Rossijskaya mul'tikonferenciya po problemam upravleniya: materialy konferencii. GNC RF OAO "KONCERN "CNII "ELEKTROPRIBOR". 2014. S. 141-146.

35.	Kefeli I.F., Malafeev O.A., O matematicheskih modelyah global'nyh geopoliticheskih processov mnogoagentnogo vzaimodejstviya, Geopolitika i bezopasnost'. 2013. № 2 (22). S. 44-57. 

36.	Kolokoltsov V.N., Malafeyev O.A. Corruption and botnent defence: a mean field game approach, International Journal of Game Theory. 2018. T. 47. № 3. S. 977-999. 

37.	Malafeyev O.A., Redinskikh N.D., Compromise solution in the problem of change state control for the material body exposed to the external medium, V sbornike: AIP Conference Proceedings. 8th Polyakhov's Reading: Proceedings of the International Scientific Conference on Mechanics. 2018. S. 080017.

38.	Malafeyev O.A., Rylow D., Kolpak E.P., Nemnyugin S.A., Awasthi A., Corruption dynamics model, V sbornike: AIP Conference Proceedings. 2017. S. 170013.

39.	Zaitseva I., Ermakova A., Shlaev D., Malafeyev O., Kolesin I., Modeling of the labour force redistribution in investment projects with account of their delay, V sbornike: IEEE International Conference on Power, Control, Signals and Instrumentation Engineering, ICPCSI 2017. 2017. S. 68-70.

40.	Malafeyev O., Onishenko V., Zubov A., Bondarenko L., Orlov V., Petrova V., Kirjanen A., Zaitseva I. Optimal location problem in the transportation network as an investment project: a numerical method.V sbornike: AIP Conference Proceedings. 2019. S. 450058. 	

41.	Vlasov M.A., Glebov V.V., Malafeyev O.A., Novichkov D.N., Experimental study of an electron beam in drift space, Soviet Journal of Communications Technology and Electronics. 1986. T. 31. № 3. S. 145.

42.	Malafeev O.A., O dinamicheskih igrah s zavisimymi dvizheniyami, Doklady Akademii nauk SSSR. 1973. T. 213. № 4. S. 783-786. 

43.	Malafeev O.A., Redinskih N.D., Alferov G.V., Model' aukciona s korrupcionnoj komponentoj, Vestnik Permskogo universiteta. Seriya: Matematika. Mekhanika. Informatika. 2015. № 1 (28). S. 30-34. 

44.	Malafeev O.A., Konfliktno upravlyaemye processy so mnogimi uchastnikami, avtoreferat dis. doktora fiziko-matematicheskih nauk / LGU im. A. A. ZHdanova. Leningrad, 1987, 44 s.

45.	Ivanyukovich G.A., Malafeyev O.A., Zaitseva I.V., Kovshov A.M., Zakharov V.V., Zakharova N.I. To the evaluation of the parameters of the regression equation between the radiometric and geological testing, V sbornike: JOP Conference Series: Metrological Support of Innovative Technologies. Krasnoyarsk Science and Technology City Hall of the Russian Union of Scientific and Engineering Associations. Krasnoyarsk, Russia, 2020. S. 32079.

46.	Malafeev O.A., Sushchestvovanie situacii ravnovesiya v beskoalicionnyh differencial'nyh igrah dvuh lic s nezavisimymi dvizheniyami, Vestnik Leningradskogo universiteta. Seriya 1: Matematika, mekhanika, astronomiya. 1980. № 4. S. 12-16. 

47.	Malafeev O.A., Stohasticheskoe ravnovesie v obshchej modeli konkurentnoj dinamiki, V sbornike: Matematicheskoe modelirovanie i prognoz social'no-ekonomicheskoj dinamiki v usloviyah konkurencii i neopredelennosti. Sbornik trudov. Sankt-Peterburg, 2004. S. 143-154.
 	
48.	Malafeev O.A., Eremin D.S., Konkurentnaya linejnaya model' ekonomiki, V sbornike: Processy upravleniya i ustojchivost'. Trudy XXXIX mezhdunarodnoj nauchnoj konferencii aspirantov i studentov. pod redakciej N. V. Smirnova, G. SH. Tamasyana. 2008. S. 425-435. 

49.	Malafeyev O.A., Redinskikh N.D., Quality estimation of the geopolitical actor development strategy, Source of the Document 2017 Constructive Nonsmooth Analysis and Related Topics (Dedicated to the Memory of V.F. Demyanov), CNSA 2017 – Proceedings, http://dx.doi.org/10.1109/CNSA.2017.7973986

50.	Ivashov L.G., Kefeli I.F., Malafeev O.A., Global'naya arkticheskaya igra i ee uchastniki, Geopolitika i bezopasnost'. 2014. № 1 (25). S. 34-49. 
51.	Kolokoltsov V.N., Malafeev O.A., Mean field game model of corruption, Stat'ya v otkrytom arhive  № 1507.03240 12.07.2015

52.	Zaitseva I.V., Malafeyev O.A., Zakharov V.V., Zakharova N.I., Orlova A.Yu. Dynamic distribution of labour resources by region of investment, V sbornike: Journal of Physics: Conference Series. 2. Ser. "II International Scientific and Practical Conference on Mathematical Modeling, Programming and Applied Mathematics, ISPCMMPAM 2020" 2020. S. 012073.

53.	Malafeev O.A.,Sushchestvovanie situacij ravnovesiya v differencial'nyh beskoalicionnyh igrah so mnogimi uchastnikami, Vestnik Leningradskogo universiteta. Seriya 1: Matematika, mekhanika, astronomiya. 1982. № 13. S. 40-46. 

54.	Malafeev O.A., Petrosyan L.A., Igra prostogo presledovaniya na ploskosti s prepyatstviem, Upravlyaemye sistemy. 1971. № 9. S. 31-42. 

55.	Malafeyev O., Galtsov M., Zaitseva I., Sakhnyuk P., Zakharov V., Kron R. Analysis of trading algorithms on the platform QIUK, V sbornike: Proceedings - 2020 2nd International Conference on Control Systems, Mathematical Modeling, Automation and Energy Efficiency, SUMMA 2020. 2. 2020. S. 305-311.

56.	Malafeev O.A., Ustojchivye beskoalicionnye igry N lic, Vestnik Leningradskogo universiteta. Seriya 1: Matematika, mekhanika, astronomiya. 1978. № 4. S. 55-58. 

57.	Malafeev O.A., CHernyh K.S., Prognosticheskaya model' zhiznennogo cikla firmy v konkurentnoj srede, V sbornike: Matematicheskoe modelirovanie i prognoz social'no-ekonomicheskoj dinamiki v usloviyah konkurencii i neopredelennosti. Sbornik trudov. Sankt-Peterburg, 2004. S. 239-255.

58.	Drozdov G.D., Malafeev O.A., Modelirovanie tamozhennogo dela. Sankt-Peterburg, 2013, Izd-vo SPbGUSE, 255 s.

59.	Pichugin YU.A., Malafeev O.A., Alferov G.V., Ocenivanie parametrov v zadachah konstruirovaniya mekhanizmov robotov-manipulyatorov, V sbornike: Ustojchivost' i processy upravleniya. Materialy III mezhdunarodnoj konferencii. 2015. S. 141-142.  

60.	Malafeev O.A., Zajceva I.V., Komarov A.A., SHvedkova T.YU., Model' korrupcionnogo vzaimodejstviya mezhdu kommercheskoj organizaciej i otdelom po bor'be s korrupciej,V knige: Linejnaya algebra s prilozheniyami k modelirovaniyu korrupcionnyh sistem i processov.  Malafeev O.A., Sotnikova N.N., Zajceva I.V., Pichugin YU.A., Kostyukov K.I., Hitrov G.M. uchebnoe posobie. Stavropol', 2016. S. 342-351. 	 

61.	Malafeev O.A., Strekopytova O.S., Teoretiko-igrovaya model' vzaimodejstviya korrumpirovannogo chinovnika s klientom: psihologicheskie aspekty, V knige: Vvedenie v modelirovanie korrupcionnyh sistem i processov.  Malafeev O.A., i dr. Kollektivnaya monografiya. Pod obshchej redakciej d.f.-m.n., professora O. A. Malafeeva. Stavropol', 2016. S. 134-151.
 
62.	Zaitseva I.V., Malafeyev O.A., Zakharov V.V., Smirnova T.E., Novozhilova L.M. Mathematical model of network flow control, V sbornike: IOP Conference Series: Materials Science and Engineering. 1. Ser. "1st International Conference on Innovative Informational and Engineering Technologies, IIET 2020" 2020. S. 012036.

63.	Malafeev O.A., Petrosyan L.A., Differencial'nye mnogokriterial'nye igry so mnogimi uchastnikami, Vestnik Leningradskogo universiteta. Seriya 1: Matematika, mekhanika, astronomiya. 1989. № 3. S. 27-31. 

64.	Malafeev O.A., Kefeli I.F., Nekotorye zadachi  obespecheniya oboronnoj bezopasnosti, Geopolitika i bezopasnost'. 2013. № 3 (23). S. 84-92. 

65.	Malafeev O.A., Redinskih N.D., Stohasticheskij analiz dinamiki korrupcionnyh gibridnyh setej, V sbornike: Ustojchivost' i kolebaniya nelinejnyh sistem upravleniya (konferenciya Pyatnickogo). Materialy XIII Mezhdunarodnoj konferencii. 2016. S. 249-251.

66.	Malafeev O.A., Farvazov K.M., Statisticheskij analiz indikatorov korrupcionnoj deyatel'nosti v sisteme gosudarstvennyh i municipal'nyh zakupok, V knige: Vvedenie v modelirovanie korrupcionnyh sistem i processov.  Malafeev O.A. i dr. Kollektivnaya monografiya. Pod obshchej redakciej d.f.-m.n., professora O. A. Malafeeva. Stavropol', 2016. S. 209-217.

67.	Malafeev O.A., Zajceva I.V., Zenovich O.S., Rumyancev N.N., Grigor'eva K.V., Ermakova A.N., Rezen'kov D.N., SHlaev D.V., Model' raspredeleniya resursov pri vzaimodejstvii korrumpirovannoj struktury s antikorrupcionnym agentstvom, V knige: Vvedenie v modelirovanie korrupcionnyh sistem i processov.  Malafeev O.A. i dr. Kollektivnaya monografiya. Pod obshchej redakciej d.f.-m.n., professora O. A. Malafeeva. Stavropol', 2016. S. 102-106. 
	
68.	Malafeyev O., Parfenov A., Smirnova T., Zubov A., Bondarenko L., Ugegov N., Strekopytova M., Strekopytov S., Zaitseva I. Game-theoretical model of cooperation between producers, V sbornike: AIP Conference Proceedings. 2019. S. 450059. 

69.	Neverova E.G., Malafeev O.A., Alferov G.V., Nelinejnaya model' upravleniya antikorrupcionnymi meropriyatiyami, V sbornike: Ustojchivost' i processy upravleniya. Materialy III mezhdunarodnoj konferencii. 2015. S. 445-446.

70.	Malafeev O.A., Marahov V.G., Evolyucionnyj mekhanizm dejstviya istochnikov i dvizhushchih sil grazhdanskogo obshchestva v sfere finansovoj i ekonomicheskoj komponenty XXI veka, V sbornike: K. Marks i budushchee filosofii Rossii.  Busov S.V., Dudnik S.I. i dr. Kollektivnaya monografiya pod redakciej zasluzhennogo deyatelya nauki RF doktora filosofskih nauk, professora V. G Marahova. Sankt-Peterburg, 2016. S. 112-135.

71.	Malafeev O.A., Borodina T.S., Kvasnoj M.A., Novozhilova L.M., Smirnov I.A., Teoretiko-igrovaya zadacha o vliyanii konkurencii v korrupcionnoj srede, V knige: Vvedenie v modelirovanie korrupcionnyh sistem i processov.  Malafeev O.A. i dr. Kollektivnaya monografiya. Pod obshchej redakciej d.f.-m.n., professora O. A. Malafeeva. Stavropol', 2016. S. 88-96.

72.	Malafeev O.A., Zajceva I.V., Koroleva O.A., Strekopytova O.S., Model' zaklyucheniya kontraktov s vozmozhno korrumpirovannym chinovnikom-principalom, V knige: Vvedenie v modelirovanie korrupcionnyh sistem i processov.  Malafeev O.A. i dr. Kollektivnaya monografiya. Pod obshchej redakciej d.f.-m.n., professora O. A. Malafeeva. Stavropol', 2016. S. 114-124.

73.	Malafeev O.A., Sajfullina D.A., Mnogoagentnoe vzaimodejstvie v transportnoj zadache s korrupcionnoj komponentoj, V knige: Vvedenie v modelirovanie korrupcionnyh sistem i processov.  Malafeev O.A. i dr. Kollektivnaya monografiya. Pod obshchej redakciej d.f.-m.n., professora O. A. Malafeeva. Stavropol', 2016. S. 218-224.

74.	Malafeev O.A., Neverova E.G., Smirnova T.E., Miroshnichenko A.N., Matematicheskaya model' processa vyyavleniya korrupcionnyh elementov v gosudarstvennoj sisteme upravleniya, V knige: Vvedenie v modelirovanie korrupcionnyh sistem i processov.  Malafeev O.A. i dr. Kollektivnaya monografiya. Pod obshchej redakciej d.f.-m.n., professora O. A. Malafeeva. Stavropol', 2016. S. 152-179.

75.	Malafeev O.A., Koroleva O.A., Neverova E.G., Model' aukciona pervoj ceny s vozmozhnoj korrupciej, V knige: Vvedenie v modelirovanie korrupcionnyh sistem i processov.  Malafeev O.A. i dr. Kollektivnaya monografiya. Pod obshchej redakciej d.f.-m.n., professora O. A. Malafeeva. Stavropol', 2016. S. 96-102. 
	
76.	Malafeev O.A., Neverova E.G., Petrov A.N., Model' processa vyyavleniya korrupcionnyh epizodov posredstvom inspektirovaniya otdelom po bor'be s korrupciej, V knige: Vvedenie v modelirovanie korrupcionnyh sistem i processov.  Malafeev O.A. i dr. Kollektivnaya monografiya. Pod obshchej redakciej d.f.-m.n., professora O. A. Malafeeva. Stavropol', 2016. S. 106-114.

77.	Malafeev O.A., Parfenov A.P., Dinamicheskaya model' mnogoagentnogo vzaimodejstviya mezhdu agentami korrupcionnoj seti, V knige: Vvedenie v modelirovanie korrupcionnyh sistem i processov.  Malafeev O.A. i dr. Kollektivnaya monografiya. Pod obshchej redakciej d.f.-m.n., professora O. A. Malafeeva. Stavropol', 2016. S. 55-63.

78.	Malafeev O.A., Andreeva M.A., Gus'kova YU.YU., Mal'ceva A.S., Poisk podvizhnogo ob"ekta pri razlichnyh informacionnyh usloviyah, V knige: Vvedenie v modelirovanie korrupcionnyh sistem i processov.  Malafeev O.A. i dr. Kollektivnaya monografiya. Pod obshchej redakciej d.f.-m.n., professora O. A. Malafeeva. Stavropol', 2016. S. 63-88.

79.	Zajceva I.V., Popova M.V., Malafeev O.A., Postanovka zadachi optimal'nogo raspredeleniya trudovyh resursov po predpriyatiyam s uchetom izmenyayushchihsya uslovij, V knige: Innovacionnaya ekonomika i promyshlennaya politika regiona (EKOPROM-2016). trudy mezhdunarodnoj nauchno-prakticheskoj konferencii. pod redakciej A.V. Babkina. 2016. S. 439-443. 	

80.	Malafeev O.A., Novozhilova L.M., Redinskih N.D., Rylov D.S., Gus'kova YU.YU., Teoretiko-igrovaya model' raspredeleniya korrupcionnogo dohoda, V knige: Vvedenie v modelirovanie korrupcionnyh sistem i processov.  Malafeev O.A. i dr. Kollektivnaya monografiya. Pod obshchej redakciej d.f.-m.n., professora O. A. Malafeeva. Stavropol', 2016. S. 125-134.

81.	Malafeev O.A., Salimov V.A., SHarlaj A.S., Algoritm ocenki bankom kreditosposobnosti klientov pri nalichii korrupcionnoj sostavlyayushchej, Vestnik Permskogo universiteta. Seriya: Matematika. Mekhanika. Informatika. 2015. № 1 (28). S. 35-38. 

82.	Kefeli I.F., Malafeev O.A., problemy ob"edineniya interesov gosudarstv EAES, SHOS i BRIKS v kontekste teorii kooperativnyh igr, Geopolitika i bezopasnost'. 2015. № 3 (31). S. 33-41. 

83.	Kefeli I.F., Malafeev O.A., Marahov V.G. i dr. Filosofskie strategii social'nyh preobrazovanij XXI veka, Pod redakciej V. G. Marahova / Sankt-Peterburg, Izd-vo SPbGU, 2014, 144 s.

84.	Malafeev O.A., Andreeva M.A., Alferov G.V., Teoretiko-igrovaya model' poiska i perekhvata v N-sektornom  regione ploskosti, Processy upravleniya i ustojchivost'. 2015. T. 2. № 1. S. 652-658. 

85.	Zajceva I.V., Malafeev O.A. Issledovanie korrupcionnyh processov i sistem matematicheskimi metodami, Innovacionnye tekhnologii v mashinostroenii, obrazovanii i ekonomike. 2017. T. 3. № 1-1 (3). S. 7-13.

86.	Kolchedancev L.M., Legalov I.N., Bad'in G.M., Malafeev O.A., Aleksandrov E.E., Gerchiu A.L., Vasil'ev YU.G. Stroitel'stvo i ekspluataciya energoeffektivnyh zdanij (teoriya i praktika s uchetom korrupcionnogo faktora) (Passivehouse), Kollektivnaya monografiya / Borovichi, 2015, 170 S.

87.	Malafeev O.A., Novozhilova L.M., Kvasnoj M.A., Legalov I.N., Primenenie metodov setevogo analiza pri proizvodstve energoeffektivnyh zdanij s uchetom korrupcionnogo faktora, V knige: Stroitel'stvo i ekspluataciya energoeffektivnyh zdanij (teoriya i praktika s uchetom korrupcionnogo faktora) (Passivehouse).  Kolchedancev L.M., Legalov I.N., Bad'in G.M., Malafeev O.A., Aleksandrov E.E., Gerchiu A.L., Vasil'ev YU.G. Kollektivnaya monografiya. Borovichi, 2015. S. 146-161. 

88.	Malafeyev O., Awasthi A., Zaitseva I., Rezenkov D., Bogdanova S., A dynamic model of functioning of a bank, V sbornike: AIP Conference Proceedings. International Conference on Electrical, Electronics, Materials and Applied Science. 2018. S. 020042. 	

89.	Malafeev O.A., Obzor literatury po modelirovaniyu korrupcionnyh sistem i processov, ch.I, V knige: Vvedenie v modelirovanie korrupcionnyh sistem i processov.  Malafeev O.A. dr. kollektivnaya monografiya. pod obshchej redakciej d.f. - m.n. , professora O. A. Malafeeva. Stavropol', 2016. S. 9-17.

90.	Asaul A.N., Lyulin P.B., Malafeev O.A., Matematicheskoe modelirovanie vzaimodejstvij organizacii kak zhivoj sistemy, Vestnik Hmel'nickogo nacional'nogo universiteta Ekonomicheskie nauki. 2013. № 6-2 (206). S. 215-220. 

91.	Awasthi A., Malafeev O.A. Is the indian dtock market efficient – a comprehensive study of Bombay stock exchange indices, Stat'ya v otkrytom arhive  № 1510.03704 10.10.2015 	

92.	Malafeev O.A., Demidova D.A., Modelirovanie processa vzaimodejstviya korrumpirovannogo predpriyatiya federal'nogo otdela po bor'be s korrupciej. V knige: Vvedenie v modelirovanie korrupcionnyh sistem i processov.  Malafeev O.A. i dr. kollektivnaya monografiya. pod obshchej redakciej d.f. - m.n. , professora O. A. Malafeeva. Stavropol', 2016. S. 140-152.

93.	Malafeev O.A., CHernyh K.S., Matematicheskoe modelirovanie razvitiya kompanii. Ekonomicheskoe vozrozhdenie Rossii. 2005, № 2. S. 23.

94.	Kulakov F.M., Alferov G.V., Malafeev O.A., Kinematicheskij analiz ispolnitel'noj sistemy manipulyacionnyh robotov, Problemy mekhaniki i upravleniya: Nelinejnye dinamicheskie sistemy. 2014. № 46. S. 31-38. 

95.	Kulakov F.M., Alferov G.V., Malafeev O.A., Dinamicheskij analiz ispolnitel'noj sistemy manipulyacionnyh robotov, Problemy mekhaniki i upravleniya: Nelinejnye dinamicheskie sistemy. 2014. № 46. S. 39-46. 

96.	Zajceva I.V., Malafeev O.A., Stepkin A.V., CHernousov M.V., Kosoblik E.V. Modelirovanie ciklichnosti razvitiya v sisteme ekonomik, Perspektivy nauki. 2020. № 10 (133). S. 173-176.

97.	Bure V.M., Malafeev O.A., Some game-theoretical models of conflict in finance, Nova Journal of Mathematics, Game Theory, and Algebra. 1996. T. 6. № 1. S. 7-14. 	

98.	Malafeev O.A., Redinskih N.D., Parfenov A.P., Smirnova T.E., Korrupciya v modelyah aukciona pervoj ceny, V sbornike: Instituty i mekhanizmy innovacionnogo razvitiya: mirovoj opty i rossijskaya praktika. Sbornik nauchnyh statej 4-j Mezhdunarodnoj nauchno-prakticheskoj konferencii. Otvetstvennyj redaktor Gorohov A.A.. 2014. S. 250-253. 
	 
99.	Marahov V.G., Malafeev O.A., Dialog filosofa i matematika: «O filosofskih aspektah matematicheskogo modelirovaniya social'nyh preobrazovanij XXI veka» V sbornike: Filosofiya poznaniya i tvorchestvo zhizni. Sbornik statej. 2014. S. 279-292.

100. Aswath Damodaran, Investment Valuation: Tools and Techniques for Determining the Value of Any Asset, 3rd Edition, 2012, Willey, 992 p.

\newpage

\section*{Appendix}

\begin{table}[h!]
\small
\begin{center}
\begin{tabular}{ | m{2.3cm} | m{1.25cm} | m{1.25cm} | m{1.25cm}| m{1.25cm} | m{1.25cm} | m{1.25cm} | } 
  \hline
 \cellcolor{lightgray}\textbf{Index} & \cellcolor{lightgray}\textbf{2020} & \cellcolor{lightgray}\textbf{2021} & \cellcolor{lightgray}\textbf{2022} & \cellcolor{lightgray}\textbf{2023} & \cellcolor{lightgray}\textbf{2024} & \cellcolor{lightgray}\textbf{Post-forecast period} \\
 \hline
\textbf{Advertising revenue} & \textbf{84 319} & \textbf{98 983} & \textbf{113 639} & \textbf{128 308} & \textbf{142 979} & \textbf{157 654} \\
 \hline
\textbf{Other revenue} & \textbf{1 249} & \textbf{1 456} & \textbf{1 672} & \textbf{1 875} & \textbf{2 074} & \textbf{2 271} \\
 \hline
\textbf{Cost price} & 10 723 & 11 332 & 11 569 & 11 434 & 10 928 & 10 051 \\
 \hline
\textbf{Research and development} & 18 050 & 21 356 & 24 712 & 28 117 & 31 573 & 35 078 \\
 \hline
\textbf{Marketing and sales} & 11 745 & 13 786 & 15 823 & 17 855 & 19 884 & 21 911 \\
 \hline
\textbf{Administrative expenses} & 6 762 & 7 641 & 8 432 & 9 136 & 9 752 & 10 281 \\
 \hline 
\textbf{Total cost and expenses} & \textbf{47 280} & \textbf{54 114} & \textbf{60 535} & \textbf{66 542} & \textbf{72 137} & \textbf{77 321} \\
 \hline
\textbf{Operating income} & \textbf{38 289} & \textbf{46 325} & \textbf{54 776} & \textbf{63 640} & \textbf{72 916} & \textbf{82 604} \\
 \hline
\textbf{Other income} & 1 004 & 1 223 & 1 495 & 1 831 & 2 248 & 2 768 \\
 \hline
\textbf{Income before tax} & 39 293 & 47 549 & 56 271 & 65 471 & 75 165 & 85 372 \\
 \hline
\textbf{Income tax} & 8 251 & 9 985 & 11 817 & 13 749 & 15 785 & 17 928 \\
 \hline
\textbf{Net income} & \textbf{31 041} & \textbf{37 563} & \textbf{44 454} & \textbf{51 722} & \textbf{59 380} & \textbf{67 444} \\
 \hline
\end{tabular}
\caption*{Table 9. Facebook: Cash Flow Forecast, millions of US dollars}
\end{center}
\end{table}

\begin{table}[h!]
\small
\begin{center}
\begin{tabular}{ | m{2.1cm} | m{1.2cm} | m{1.2cm} | m{1.2cm}| m{1.2cm} | m{1.2cm} | m{1.2cm} | } 
  \hline
 \cellcolor{lightgray}\textbf{Index} & \cellcolor{lightgray}\textbf{2020} & \cellcolor{lightgray}\textbf{2021} & \cellcolor{lightgray}\textbf{2022} & \cellcolor{lightgray}\textbf{2023} & \cellcolor{lightgray}\textbf{2024} & \cellcolor{lightgray}\textbf{Post-forecast period} \\
 \hline
\textbf{EBIT} & 39 293 & 47 549 & 56 271 & 65 471 & 75 165 & 85 372 \\
 \hline
\textbf{Income tax} & (8 251) & (9 985) & (11 817) & (13 749) & (15 785) & (17 928) \\
 \hline
\textbf{Depreciation and Amortization} & 5 925 & 5 878 & 5 619 & 5 190 & 4 634 & 4 241 \\
 \hline
\textbf{CaPex} & (20 023) & (25 512) & (30 442) & (35 670) & (41 195) & (44 299) \\
 \hline
\textbf{$\bigtriangleup$ NWC} & (8 256) & (1 220) & 1 477 & 4 175 & 6 872 & 9 569 \\
 \hline
\textbf{FCFF} & \textbf{8 687} & \textbf{16 710} & \textbf{21 108} & \textbf{25 417} & \textbf{29 690} & \textbf{36 955} \\
 \hline 
\end{tabular}
\caption*{Table 10. Facebook: FCFF values, millions of US dollars}
\end{center}
\end{table}

\begin{table}[h!]
\small
\begin{center}
\begin{tabular}{ | m{2.3cm} | m{1.5cm} | m{1.5cm} | m{1.5cm}| m{1.5cm} | m{1.5cm} | m{1.5cm} | } 
  \hline
 \cellcolor{lightgray}\textbf{Index} & \cellcolor{lightgray}\textbf{2020} & \cellcolor{lightgray}\textbf{2021} & \cellcolor{lightgray}\textbf{2022} & \cellcolor{lightgray}\textbf{2023} & \cellcolor{lightgray}\textbf{2024} & \cellcolor{lightgray}\textbf{Post-forecast period} \\
 \hline
\textbf{Advertising revenue} & \textbf{3 368 695} & \textbf{3 723 268} & \textbf{4 106 500} & \textbf{4 481 950} & \textbf{4 859 079} & \textbf{5 237 886} \\
 \hline
\textbf{Data licensing and other revenue} & \textbf{523 051} & \textbf{577 111} & \textbf{613 615} & \textbf{657 901} & \textbf{700 508} & \textbf{741 436} \\
 \hline
\textbf{Cost price} & 1 201 450 & 1 285 615 & 1 365 011 & 1 436 211 & 1 499 215 & 1 554 022 \\
 \hline
\textbf{Research and development} & 736 870 & 797 508 & 858 146 & 918 784 & 979 422 & 1 040 061 \\
 \hline
\textbf{Marketing and sales} & 1 009 566 & 1 032 671 & 1 042 473 & 1 036 093 & 1 013 531 & 974 785 \\
 \hline
\textbf{Administrative expenses} & 413 018 & 451 592 & 490 167 & 528 742 & 567 316 & 605 891 \\
 \hline 
\textbf{Total cost and expenses} & \textbf{3 360 903} & \textbf{3 567 386} & \textbf{3 755 798} & \textbf{3 919 830} & \textbf{4 059 484} & \textbf{4 174 759} \\
 \hline
\textbf{Operating income} & \textbf{530 842} & \textbf{732 992} & \textbf{964 317} & \textbf{1 220 020} & \textbf{1 500 102} & \textbf{1 804 564} \\
 \hline
\textbf{Interest income (expense), net} & 15 587 & 11 650 & 7 714 & 3 777 & (159) & (4 096) \\
 \hline
\textbf{Other income (expense), net} & 24 136 & 23 790 & 23 444 & 23 098 & 22 752 & 22 407 \\
 \hline
\textbf{Income before tax} & 570 564 & 768 432 & 995 474 & 1 246 895 & 1 522 695 & 1 822 874 \\
 \hline
\textbf{Income tax} & 119 818 & 161 371 & 209 050 & 261 848 & 319 766 & 382 804 \\
 \hline
\textbf{Net income} & \textbf{450 746} & \textbf{607 061} & \textbf{786 424} & \textbf{985 047} & \textbf{1 202 929} & \textbf{1 440 071} \\
 \hline
\end{tabular}
\caption*{Table 11. Twitter: Cash Flow Forecast, thousands of US dollars}
\end{center}
\end{table}

\begin{table}[h!]
\small
\begin{center}
\begin{tabular}{ | m{2.1cm} | m{1.4cm} | m{1.4cm} | m{1.4cm}| m{1.4cm} | m{1.49cm} | m{1.4cm} | } 
  \hline
 \cellcolor{lightgray}\textbf{Index} & \cellcolor{lightgray}\textbf{2020} & \cellcolor{lightgray}\textbf{2021} & \cellcolor{lightgray}\textbf{2022} & \cellcolor{lightgray}\textbf{2023} & \cellcolor{lightgray}\textbf{2024} & \cellcolor{lightgray}\textbf{Post-forecast period} \\
 \hline
\textbf{EBIT} & 586 151 & 780 082 & 1 003 187 & 1 250 672 & 1 522 536 & 1 818 779 \\
 \hline
\textbf{Income tax} & (119 818) & (161 371) & (209 050) & (261 848) & (319 766) & (382 804) \\
 \hline
\textbf{Depreciation and Amortization} & 527 655 & 587 422 & 647 189 & 706 957 & 766 724 & 826 491 \\
 \hline
\textbf{CaPex} & (550 873) & (610 761) & (670 649) & (730 537) & (790 424) & (850 312) \\
 \hline
\textbf{$\bigtriangleup$ NWC} & (69 864) & (363 185) & (313 089) & (243 429) & (173 769) & (418 803) \\
 \hline
\textbf{FCFF} & \textbf{373 250} & \textbf{232 187} & \textbf{457 590} & \textbf{721 815} & \textbf{1 005 300} & \textbf{993 351} \\
 \hline 
\end{tabular}
\caption*{Table 12. Twitter: FCFF values, thousands of US dollars}
\end{center}
\end{table}

\begin{table}[h!]
\small
\begin{center}
\begin{tabular}{ | m{2.3cm} | m{1.49cm} | m{1.49cm} | m{1.49cm}| m{1.49cm} | m{1.49cm} | m{1.49cm} | } 
  \hline
 \cellcolor{lightgray}\textbf{Index} & \cellcolor{lightgray}\textbf{2020} & \cellcolor{lightgray}\textbf{2021} & \cellcolor{lightgray}\textbf{2022} & \cellcolor{lightgray}\textbf{2023} & \cellcolor{lightgray}\textbf{2024} & \cellcolor{lightgray}\textbf{Post-forecast period} \\
 \hline
\textbf{Advertising revenue} & \textbf{2 222 290} & \textbf{2 393 648} & \textbf{2 564 735} & \textbf{2 735 574} & \textbf{2 906 184} & \textbf{3 076 583} \\
 \hline
\textbf{Other revenue} & \textbf{483 468} & \textbf{512 887} & \textbf{542 223} & \textbf{571 482} & \textbf{600 671} & \textbf{629 794} \\
 \hline
\textbf{Cost price (advertising)} & 417 146 & 443 276 & 469 407 & 495 539 & 521 674 & 547 810 \\
 \hline
\textbf{Cost price (other)} & 170 695 & 179 676 & 188 626 & 197 548 & 206 444 & 215 316 \\
 \hline
\textbf{Marketing and sales} & 643 152 & 700 155 & 757 158 & 814 161 & 871 164 & 928 167 \\
 \hline
\textbf{Product development} & 449 013 & 442 742 & 477 534 & 512 326 & 547 118 & 581 910 \\
 \hline 
\textbf{Administrative expenses} & 317 124 & 335 808 & 354 491 & 373 175 & 391 858 & 410 542 \\
 \hline 
\textbf{Total cost and expenses} & \textbf{1 997 131} & \textbf{2 101 657} & \textbf{2 247 216} & \textbf{2 392 749} & \textbf{2 538 258} & \textbf{2 683 745} \\
 \hline
\textbf{Operating income} & \textbf{708 627} & \textbf{804 879} & \textbf{859 743} & \textbf{914 307} & \textbf{968 597} & \textbf{1 022 633} \\
 \hline
\textbf{Interest and other income, net} & 46 311 & 51 358 & 56 418 & 61 490 & 66 573 & 71 665 \\
 \hline
\textbf{Investments} & 48 721 & 44 555 & 40 388 & 36 222 & 32 056 & 27 890 \\
 \hline
\textbf{Changes in fair value due to return on investment, net} & 181 825 & 200 007 & 220 008 & 242 008 & 266 209 & 292 830 \\
 \hline
\textbf{Impairment of investments} & (412 382) & (452 648) & (492 808) & (532 871) & (572 845) & (612 737) \\
 \hline
\textbf{Income before tax} & 573 100 & 648 151 & 683 749 & 721 157 & 760 589 & 802 280 \\
 \hline
\textbf{Income tax} & (143 275) & (162 038) & (170 937) & (180 289) & (190 147) & (200 570) \\
 \hline
\textbf{Net income} & \textbf{429 825} & \textbf{486 113} & \textbf{512 812} & \textbf{540 868} & \textbf{570 442} & \textbf{601 710} \\
 \hline
\end{tabular}
\caption*{Table 13. Sina Weibo: Cash Flow Forecast, thousands of US dollars}
\end{center}
\end{table}

\begin{table}[h!]
\small
\begin{center}
\begin{tabular}{ | m{2.1cm} | m{1.78cm} | m{1.4cm} | m{1.4cm}| m{1.4cm} | m{1.4cm} | m{1.4cm} | } 
  \hline
 \cellcolor{lightgray}\textbf{Index} & \cellcolor{lightgray}\textbf{2020} & \cellcolor{lightgray}\textbf{2021} & \cellcolor{lightgray}\textbf{2022} & \cellcolor{lightgray}\textbf{2023} & \cellcolor{lightgray}\textbf{2024} & \cellcolor{lightgray}\textbf{Post-forecast period} \\
 \hline
\textbf{EBIT} & 577 731 & 653 287 & 689 391 & 727 306 & 767 247 & 809 447 \\
 \hline
\textbf{Income tax} & (143 275) & (162 038) & (170 937) & (180 289) & (190 147) & (200 570) \\
 \hline
\textbf{Depreciation and Amortization} & 73 647 & 77 275 & 83 048 & 88 821 & 94 595 & 100 368 \\
 \hline
\textbf{CaPex} & (194 652) & (209 096) & (223 514) & (237 909) & (252 282) & (266 636) \\
 \hline
\textbf{$\bigtriangleup$ NWC} & (2 013 507) & (364 056) & (363 411) & (362 822) & (362 280) & (361 780) \\
 \hline
\textbf{FCFF} & \textbf{(1 700 055)} & \textbf{(4 628)} & \textbf{14 576} & \textbf{35 107} & \textbf{57 132} & \textbf{80 829} \\
 \hline 
\end{tabular}
\caption*{Table 14. Sina Weibo: FCFF values, thousands of US dollars}
\end{center}
\end{table}

\begin{table}[h!]
\small
\begin{center}
\begin{tabular}{ | m{2.29cm} | m{1.7cm} | m{1.7cm} | m{1.7cm}| m{1.7cm} | m{1.7cm} | m{1.7cm} | } 
  \hline
 \cellcolor{lightgray}\textbf{Index} & \cellcolor{lightgray}\textbf{2020} & \cellcolor{lightgray}\textbf{2021} & \cellcolor{lightgray}\textbf{2022} & \cellcolor{lightgray}\textbf{2023} & \cellcolor{lightgray}\textbf{2024} & \cellcolor{lightgray}\textbf{Post-forecast period} \\
 \hline
\textbf{Revenue} & \textbf{25 488 935} & \textbf{31 965 571} & \textbf{38 926 079} & \textbf{46 304 752} & \textbf{54 055 191} & \textbf{62 143 297} \\
 \hline
\textbf{Cost of sales} & (9 654 924) & (10 937 862) & (12 220 801) & (13 503 739) & (14 786 678) & (16 069 616) \\
 \hline
\textbf{Gross profit (loss)} & \textbf{15 834 011} & \textbf{21 027 709} & \textbf{26 705 278} & \textbf{32 801 013} & \textbf{39 268 514} & \textbf{46 073 681} \\
 \hline
\textbf{Selling expenses} & (247 478) & (287 958) & (326 153) & (361 463) & (393 605) & (422 589) \\
 \hline 
\textbf{Administrative expenses} & (6 438 540) & (6 383 026) & (6 327 809) & (6 272 888) & (6 218 264) & (6 163 935) \\
 \hline 
\textbf{Income (loss) from sales} & \textbf{9 147 992} & \textbf{14 356 725} & \textbf{20 051 317} & \textbf{26 166 662} & \textbf{32 656 645} & \textbf{39 487 157} \\
 \hline
\textbf{Interest receivable} & 279 912 & 319 679 & 359 445 & 399 212 & 438 979 & 478 745 \\
 \hline
\textbf{Interest payable} & - & - & - & - & - & - \\
 \hline
\textbf{Other income} & 517 726 & 564 702 & 611 679 & 658 656 & 705 633 & 752 609 \\
 \hline
\textbf{Other expenses} & (779 862) & (527 739) & (912 339) & (660 216) & (1 044 816) & (792 693) \\
 \hline
\textbf{Income before tax} & \textbf{9 165 768} & \textbf{14 713 367} & \textbf{20 110 103} & \textbf{26 564 314} & \textbf{32 756 440} & \textbf{39 925 819} \\
 \hline
\textbf{Income tax} & 1 833 154 & 2 942 673 & 4 022 021 & 5 312 863 & 6 551 288 & 7 985 164 \\
 \hline
\textbf{Net income} & \textbf{7 332 614} & \textbf{11 770 694} & \textbf{16 088 082} & \textbf{21 251 451} & \textbf{26 205 152} & \textbf{31 940 655} \\
 \hline
\end{tabular}
\caption*{Table 15. VKontakte: Cash Flow Forecast, thousands of US dollars}
\end{center}
\end{table}

\begin{table}[h!]
\small
\begin{center}
\begin{tabular}{ | m{2.1cm} | m{1.6cm} | m{1.68cm} | m{1.68cm}| m{1.68cm} | m{1.68cm} | m{1.68cm} | } 
  \hline
 \cellcolor{lightgray}\textbf{Index} & \cellcolor{lightgray}\textbf{2020} & \cellcolor{lightgray}\textbf{2021} & \cellcolor{lightgray}\textbf{2022} & \cellcolor{lightgray}\textbf{2023} & \cellcolor{lightgray}\textbf{2024} & \cellcolor{lightgray}\textbf{Post-forecast period} \\
 \hline
\textbf{EBIT} & 9 445 680 & 15 033 046 & 20 469 548 & 26 963 526 & 33 195 419 & 40 404 564 \\
 \hline
\textbf{Income tax} & 1 833 154 & 2 942 673 & 4 022 021 & 5 312 863 & 6 551 288 & 7 985 164 \\
 \hline
\textbf{Depreciation and Amortization} & - & - & - & - & - & - \\
 \hline
\textbf{CaPex} & (2 281 432) & (3 063 182) & (3 976 234) & (5 022 635) & (6 204 989) & (7 526 216) \\
 \hline
\textbf{$\bigtriangleup$ NWC} & - & - & - & - & - & - \\
 \hline
\textbf{FCFF} & \textbf{8 997 402} & \textbf{14 912 537} & \textbf{20 515 335} & \textbf{27 253 753} & \textbf{33 541 718} & \textbf{40 863 512} \\
 \hline 
\end{tabular}
\caption*{Table 16. VKontakte: FCFF values, thousands of US dollars}
\end{center}
\end{table}

\begin{table}[h!]
\small
\begin{center}
\begin{tabular}{ | m{1.9cm} | m{1.6cm} | m{1.3cm} | m{1.6cm}| m{1.6cm} | m{1.2cm} | m{1.75cm} | } 
  \hline
 \cellcolor{lightgray}\textbf{Index} & \cellcolor{lightgray}\textbf{Facebook} & \cellcolor{lightgray}\textbf{Twitter} & \cellcolor{lightgray}\textbf{Pinterest} & \cellcolor{lightgray}\textbf{Snapchat} & \cellcolor{lightgray}\textbf{Sina Weibo} & \cellcolor{lightgray}\textbf{VKontakte} \\
 \hline
\textbf{Revenue} & 70 697 & 3 459 & 1 143 & 1 716 & 2 163 & 299 \\
 \hline
\textbf{EBIT} & 24 812 & 390 & (1 361) & (1 033) & 255 & 132 \\
 \hline
\textbf{EBITDA} & 30 553 & 856 & (1 333) & (946) & 316 & 132 \\
 \hline
\textbf{NI} & 18 485 & 1 466 & (1 361) & (1 034) & 109 & 109 \\
 \hline
\textbf{DAU} & 1 657 & 152 & 250 & 218 & 222 & 23 \\
 \hline
\textbf{MAU} & 2 498 & 330 & 335 & 293 & 516 & 72 \\
 \hline
\end{tabular}
\caption*{Table 17. Multiplier method: company data for calculating multipliers}
\end{center}
\end{table}

\begin{table}[h!]
\small
\begin{center}
\begin{tabular}{ | m{2.1cm} | m{1.6cm} | m{1.3cm} | m{1.6cm}| m{1.6cm} | m{1.2cm} | m{1.75cm} | } 
  \hline
 \cellcolor{lightgray}\textbf{Index} & \cellcolor{lightgray}\textbf{Facebook} & \cellcolor{lightgray}\textbf{Twitter} & \cellcolor{lightgray}\textbf{Pinterest} & \cellcolor{lightgray}\textbf{Snapchat} & \cellcolor{lightgray}\textbf{Sina Weibo} & \cellcolor{lightgray}\textbf{VKontakte} \\
 \hline
\textbf{EV/EBIT} & 17,31 & 10,70 & 23,45 & 26,71 & 20,31 & 11,60 \\
 \hline
\textbf{EV/EBITDA} & 10,36 & 9,03 & 15,65 & 19,31 & 12,81 & 3,77 \\
 \hline
\textbf{EV/R} & 10,59 & 10,97 & 10,54 & 9,42 & 12,05 & 7,66 \\
 \hline
\textbf{P/DAU} & 120,61 & 162,10 & 183,35 & 168,90 & 189,19 & 131,56 \\
 \hline
\textbf{P/MAU} & 55,72 & 89,13 & 96,69 & 85,96 & 101,66 & 83,37 \\
 \hline
\end{tabular}
\caption*{Table 18. Multiplier method: table of average values of multipliers for calculating the values for each company}
\end{center}
\end{table}

\end{document}